\documentclass[10pt,journal,compsoc]{IEEEtran}
\usepackage{color,soul} % for highlighting
\usepackage{graphicx}
\usepackage{array}
\usepackage{subfigure}
\usepackage{multirow}

\usepackage{tabularx,booktabs,enumitem}
\newcolumntype{C}{>{\centering\arraybackslash}X} % centered version of "X" type
\newcommand{\ro}[1]{\rotatebox{60}{#1}}

\newcommand{\revc}[1]{{#1}}
\newcommand{\revn}[1]{{#1}}

\ifCLASSINFOpdf
\else
\fi
\begin{document}

\title{A Review of Surface Haptics:\\ Enabling Tactile Effects on Touch Surfaces}

%
%
% author names and IEEE memberships
% note positions of commas and nonbreaking spaces ( ~ ) LaTeX will not break
% a structure at a ~ so this keeps an author's name from being broken across
% two lines.
% use \thanks{} to gain access to the first footnote area
% a separate \thanks must be used for each paragraph as LaTeX2e's \thanks
% was not built to handle multiple paragraphs
%

\author{Cagatay~Basdogan,~\IEEEmembership{Member,~IEEE,}
        Frederic~Giraud,~\IEEEmembership{Member,~IEEE,}
        
        Vincent~Levesque,~\IEEEmembership{Member,~IEEE,}
        Seungmoon Choi,~\IEEEmembership{Senior Member,~IEEE}% <-this % stops a space
\thanks{C. Basdogan is with the College
of Engineering, Koc University, Istanbul, 34450, Turkey, e-mail: cbasdogan@ku.edu.tr.}
\thanks{F. Giraud is with the University of Lille, Lille F59000, France, e-mail: frederic.giraud@univ-lille.fr.}
\thanks{V. Levesque is with the Department of Software and IT Engineering, École de Technologie Supérieure, Montreal, Canada, e-mail: vincent.levesque@etsmtl.ca.}
\thanks{S. Choi is with the Department of Computer Science and Engineering, Pohang University of Science and Technology (POSTECH), Pohang, Republic of Korea, e-mail: choism@postech.ac.kr.}% <-this % stops a space
% <-this % stops a space
}

% note the % following the last \IEEEmembership and also \thanks - 
% these prevent an unwanted space from occurring between the last author name
% and the end of the author line. i.e., if you had this:
% 
% \author{....lastname \thanks{...} \thanks{...} }
%                     ^------------^------------^----Do not want these spaces!
%
% a space would be appended to the last name and could cause every name on that
% line to be shifted left slightly. This is one of those "LaTeX things". For
% instance, "\textbf{A} \textbf{B}" will typeset as "A B" not "AB". To get
% "AB" then you have to do: "\textbf{A}\textbf{B}"
% \thanks is no different in this regard, so shield the last } of each \thanks
% that ends a line with a % and do not let a space in before the next \thanks.
% Spaces after \IEEEmembership other than the last one are OK (and needed) as
% you are supposed to have spaces between the names. For what it is worth,
% this is a minor point as most people would not even notice if the said evil
% space somehow managed to creep in.

% The paper headers
\markboth{IEEE Transactions on Haptics,~Vol.~XX, No.~XX, XX~2020}%
{Shell \MakeLowercase{\textit{et al.}}: Bare Demo of IEEEtran.cls for IEEE Journals}
% The only time the second header will appear is for the odd numbered pages
% after the title page when using the twoside option.
% 
% *** Note that you probably will NOT want to include the author's ***
% *** name in the headers of peer review papers.                   ***
% You can use \ifCLASSOPTIONpeerreview for conditional compilation here if
% you desire.

% If you want to put a publisher's ID mark on the page you can do it like
% this:
%\IEEEpubid{0000--0000/00\$00.00~\copyright~2015 IEEE}
% Remember, if you use this you must call \IEEEpubidadjcol in the second
% column for its text to clear the IEEEpubid mark.

% use for special paper notices
%\IEEEspecialpapernotice{(Invited Paper)}

% make the title area

\IEEEtitleabstractindextext{
\begin{abstract}
We review the current technology underlying surface haptics that converts passive touch surfaces to active ones (machine haptics), our perception of tactile stimuli displayed through active touch surfaces (human haptics), their potential applications (human-machine interaction), and finally the challenges ahead of us in making them available through commercial systems. This review primarily covers the tactile interactions of human fingers or hands with surface-haptics displays by focusing on the three most popular actuation methods: vibrotactile, electrostatic, and ultrasonic.
\end{abstract}

% Note that keywords are not normally used for peerreview papers.
\begin{IEEEkeywords}
surface haptics, tactile feedback, vibrotactile, electrovibration, ultrasonic, friction modulation, review, state of the art.
\end{IEEEkeywords}}

\maketitle

% For peer review papers, you can put extra information on the cover
% page as needed:
% \ifCLASSOPTIONpeerreview
% \begin{center} \bfseries EDICS Category: 3-BBND \end{center}
% \fi
%
% For peerreview papers, this IEEEtran command inserts a page break and
% creates the second title. It will be ignored for other modes.
\IEEEpeerreviewmaketitle

\section{Introduction}

\IEEEPARstart{H}{aptics} for interactive touch surfaces, also known as surface haptics, is a new area of research in the field of haptics. The goal of surface haptics is to generate tactile effects on touch surfaces such as those used in mobile phones, tablets, kiosks and information displays, and front panels of new generation home appliances and cars.

\begin{figure}[ht]
\centering
\includegraphics[width=0.99\columnwidth]{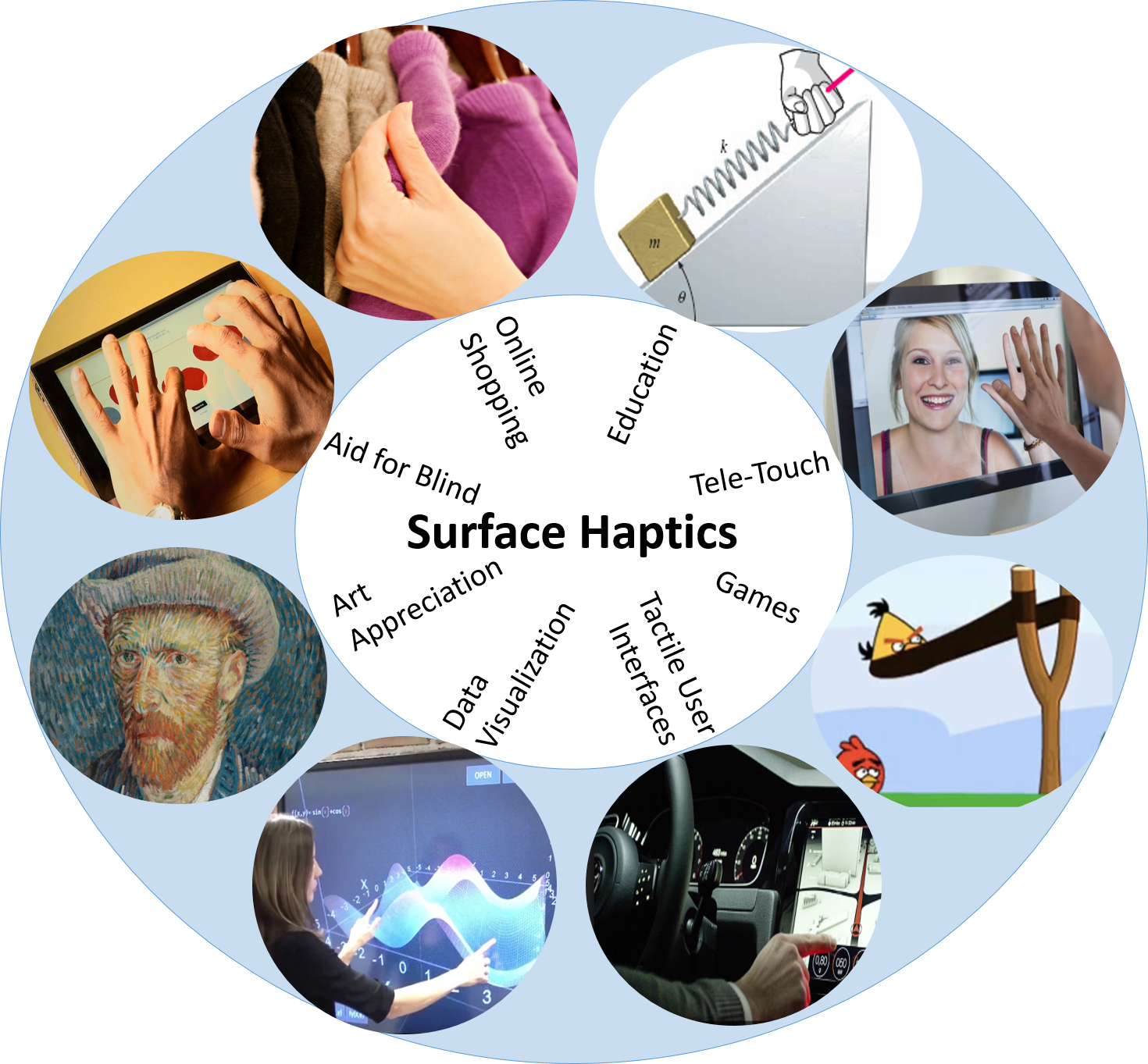}
\caption{Applications of surface haptics span many areas including online shopping, education, tele-touch, games and entertainment, tactile user interfaces, data visualization, art appreciation, and aid for blind and visually impaired}
\label{fig:applications}
\end{figure}

Integration of haptics into touch surfaces will result in new applications in user interface design, online shopping, gaming and entertainment, education, arts, and more (see Fig.~\ref{fig:applications}). \revc{Imagine that you can feel the fabric of clothes as you purchase them online on a tablet, that you can feel the edges of buttons on your mobile phone as you dial without looking, that you can play chess on an interactive touch surface and feel a frictional resistance when you make an incorrect move, that you can feel the detents of a digital knob on the touch surface of your car as you rotate it, that a visually impaired person navigates in a building using a haptic map on a tablet, or that your children feel an exotic animal on an interactive touchscreen in a classroom.} As obvious from the examples above, surface haptics is a new and exciting area of research and the number of potential applications of this technology is countless. 

This review primarily covers the tactile interactions of human \revc{fingers or hands} with surface-haptics displays by focusing on the three most popular actuation methods: vibrotactile, electrostatic, and ultrasonic. We intentionally excluded some other less-common actuation methods such as electromagnetic actuation (e.g., \cite{Jansen10}) and methods utilizing fluidic pressure \cite{Craig12} as well as kinesthetic haptic interactions (\revc{that do} not primarily involve \revc{the} tactile channel) with shape-changing surfaces (e.g., \cite{Follmer13}), indirect haptic interactions with touch surfaces such as mid-air haptics via acoustic radiation pressure (e.g., \cite{Hoshi10}) and through hand-held devices such as pen/stylus from our review (e.g., \cite{Kyung09}).

We review the current state of the art in terms of 1) machine haptics: technologies to generate tactile stimuli on touch surfaces and the physics behind \revc{them}, 2) human haptics: our tactile perception of those stimuli, 3) human-machine haptics: interactions between \revc{humans and surface haptics technologies}, leading to rendering of virtual shapes, textures, and controllers with potential applications in the areas discussed above.
%%%%%%%%%%%%%%%%%%%%%%%%%%%%%%%%%%%%%%%%%%%%%%%%%%%
\section{Machine haptics}

Here, we provide an introduction to the current actuation technologies enabling tactile feedback on touch surfaces. We first group them based on the direction of stimulation and then briefly discuss how each technology works. 

\subsection{Classification}

\begin{figure}[ht]
\centering
\includegraphics[width=1.0\columnwidth]{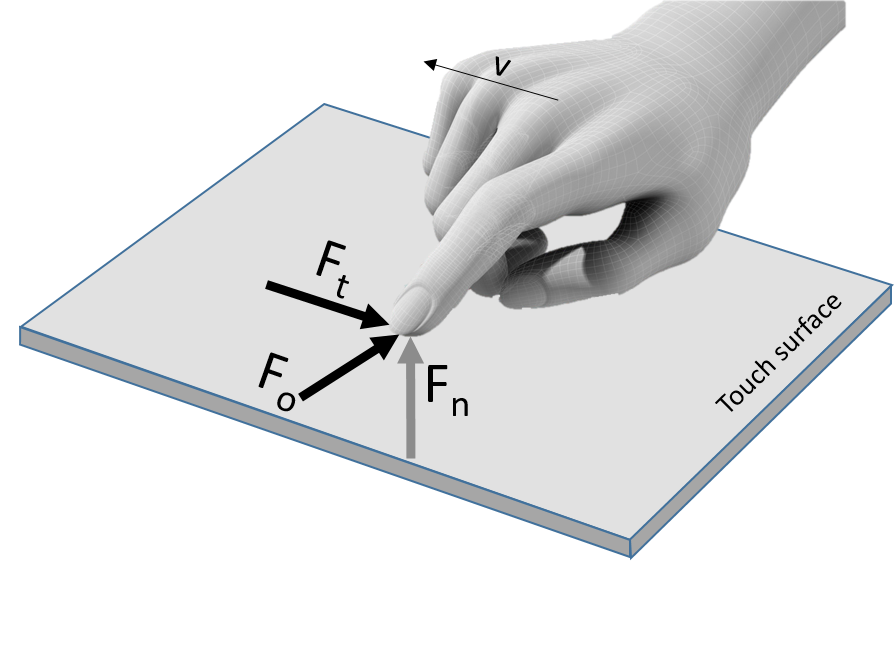}
\caption{The components of the force acting on the finger during haptic interactions with a touch surface}
\label{fig:normal_vs_tangential}
\end{figure}

\begin{figure*}[!ht]
    \begin{center}
        \setlength{\unitlength}{0.14in}% selecting unit length
\centering% used for centering Figure
\begin{picture}(53,17)  % picture environment with the size (dimensions)  % 32 length units wide, and 15 units high.
\put(19,16){
    \makebox(0,0){\large{\textbf{\textsf{Surface Haptics}}}}
    \linethickness{0.5mm}
    \put(0,-2){\line(0,1){1}\line(-1,0){11.5}\line(0,-1){1}}
    \put(0,-2){\line(0,1){1}\line(1,0){11.5}\line(0,-1){1}}
}
\put(7.5,12){\makebox(0,0){\textbf{\textsf{Force modulation}}}}
\put(30.5,12){\makebox(0,0){\textbf{\textsf{Force modulation}}}}

\put(7.5,10.5){
    \makebox(0,0){\textbf{\textsf{in normal direction}}}
    \put(0,-4){
        \linethickness{0.3mm}
        \put(0,2){\line(0,1){1}\line(-1,0){4}\line(0,-1){1}}
        \put(0,2){\line(0,1){1}\line(1,0){4}\line(0,-1){1}}
        \put(-4,0){
            \makebox(0,0){\textsf{Normal vibration}}
            %\put(0,-1.2){\makebox(0,0){\cite{Poupyrev:2002, Yao:2010, Hudin:2018,  Dhiab:2019, Emgin19}}}
        }
        \put(4,0){
            \makebox(0,0){\textsf{Pulses}}
            %\put(0,-1.2){\makebox(0,0){\cite{Bai:2011, Woo:2015, Hudin:2015, Enferad:2018}}}
        }
    }
}

\put(30.5,10.5){
    \makebox(0,0){\textbf{\textsf{in tangential plane}}}
    \put(0,-4){
        \linethickness{0.3mm}
        \put(0,2){\line(0,1){1}\line(-1,0){11}\line(0,-1){1}}
        \put(-11,0){
            \makebox(0,0){\textsf{Lateral vibration}}
            %\put(0,-1,2){\makebox(0,0){\cite{wiertlewski:2011}}}
        }
        \put(0,2){\line(0,-1){1}}
        \put(0,0){
            \makebox(0,0){\textsf{Friction modulation}}
            \put(0,-4){
                \put(-4,0){\makebox(0,0){\textsf{Electrostatic}}}
                %\put(-4,-1.2){\makebox(0,0){\cite{}}}
                \put(4,0){\makebox(0,0){\textsf{Ultrasonic}}}
                \put(4,-1){\makebox(0,0){\textsf{waves}}}
                %\put(4,-2.2){\makebox(0,0){\cite{Nara:2001, Biet:2007, %Winfield:2007, ghenna:2017}}}
            }
            \put(0,-2){\line(0,1){1}\line(-1,0){4}\line(0,-1){1}}
            \put(0,-2){\line(0,1){1}\line(1,0){4}\line(0,-1){1}}
        }
        \put(0,2){\line(0,1){1}\line(1,0){13.5}\line(0,-1){1}}
        \put(13.5,-0.2){
            \makebox(0,0){\textsf{Net tangential force}}
            \put(-4,-4){
                \makebox(0,0){\textsf{Driving}}
                \put(0,-1){\makebox(0,0){\textsf{force}}}
            }
            \put(4,-4){
                \makebox(0,0){\textsf{Asymmetric}}
                \put(0,-1){\makebox(0,0){\textsf{friction}}}
            }
            \put(0,-2){\line(0,1){1}\line(-1,0){4}\line(0,-1){1}}
            \put(0,-2){\line(0,1){1}\line(1,0){4}\line(0,-1){1}}
        }
        
    }
}

\end{picture}
    \end{center}
\caption{Classification of the current technologies for surface haptics displays based on the stimulation direction and method.}
\label{fig-clas_Fig} 
\end{figure*}
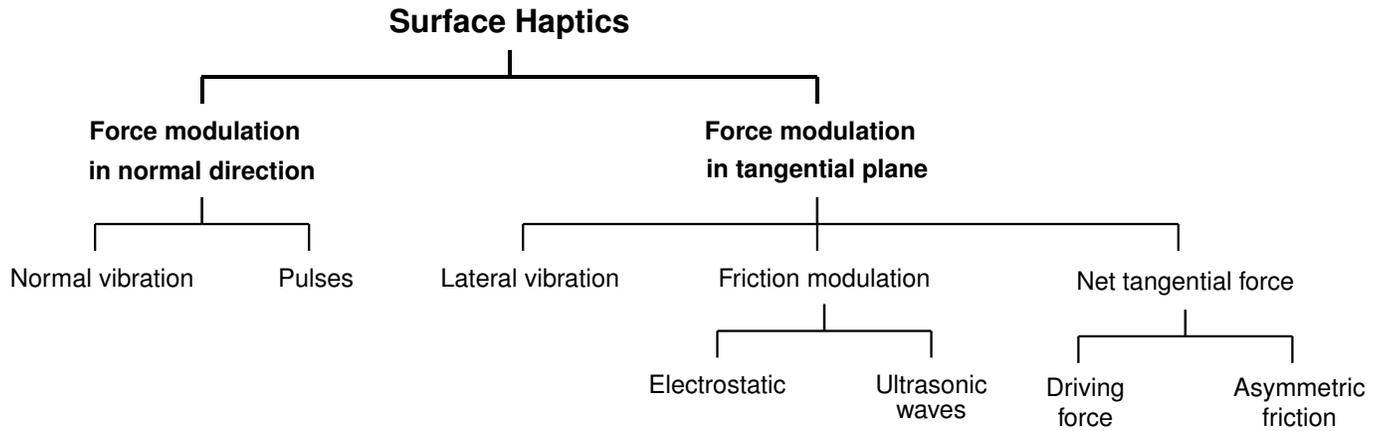

From the technological point of view, the goal of surface haptics is to design and develop new devices in order to display tactile feedback to the users by modulating the interaction forces between the finger and the touch surface. In general, the current actuation technologies can be grouped based on the direction \revc{in which} the finger is stimulated by the interaction forces\revc{,} as depicted in Fig.~\ref{fig:normal_vs_tangential}: a) normal to the surface ($F_n$), b) tangential to the surface ($F_t$ and $F_o$). In Fig. \ref{fig-clas_Fig}, we present our approach of further classifying the current technologies based on the general force decomposition given in Fig.~\ref{fig:normal_vs_tangential}.

%The first requirement is that the tactile feedback should not change the functionality of the surface. For instance, if this surface lies on top of a display, it should remain transparent. Also, user’s fingertips moves freely on the surface, and by principle, no actuator should be attached to the fingertip to produce the stimulation. Having this in mind, and apart of some exotic examples~\cite{Craig12}, it has been easier on a technological point of view not to change to geometry of the surface, but instead, to modulate the contact forces between the surface and the fingertip. To achieve such a goal, the examples proposed in the literature review can be shared into two principles: forces generated normal to the surface, or instead parallel to it.
%

When the stimulation is in the normal direction, an actuator placed in the periphery of the surface creates a mechanical vibration that propagates inside the material and reaches the finger ~\cite{Yao:2010}. This type of stimulation, denoted by \emph{normal vibration}, is directly detectable by our tactile system if it occurs at a frequency below 1 kHz. In fact, vibration actuators are already embedded in mobile phones today for this purpose and \revc{used} to alert the user about incoming calls and to provide confirmation for button press events. Although it is highly difficult to generate complex tactile effects using these simple actuators, they have been 
%preferred by the manufacturers over voice coils and piezoelectric actuators
used by manufacturers due to their low-cost and low energy requirements. 

Using multiple actuators allows for more sophisticated stimulation techniques in the normal direction that lead to unique tactile effects. For example, imposing multiple vibrations on the surface with adequate amplitude or phase modulation provides a vibration sensation midway between the actual stimulated points, which can either be static or moving. This so-called \emph{tactile phantom sensation} has been already studied intensively and \revc{used} for many applications %\cite{KimSY2009, Kim2012, Kang2012, Seo2010, Seo2013, Seo2015_WHC, Zhao2015_WHC}
~\cite{Zhao2015_WHC}; see~\cite{Park2018} for \revc{a} review. \revn{We further discuss the application of this concept to surface haptics in Section \ref{Phantom Sensations}}. Other techniques are also possible mostly to localize vibrations to a small region on the surface by obtaining constructive and destructive waves where needed \revn{(see Section \ref{localization})}. The theory behind this principle has been investigated under the names of \textit{inverse filtering}~\cite{Hudin:2018} and \textit{modal composition}~\cite{Woo:2015, Enferad:2018}.

Instead of displaying continuous vibrations, short vibration \emph{pulses} can be used to generate tactile effects in the normal direction, which is more challenging due to the resonating nature of \revc{the} touch surface. The mechanical excitation of the surface produces echoes, which have to be cancelled out in order to obtain a short burst of normal displacement at a specific location. For that purpose, multiple actuators are placed on the surface, and their control signals are synchronized in order to create constructive and destructive interference. The references for these signals can be deduced from the \textit{time reversal} theory, as in ~\cite{Bai:2011, Hudin:2013}.

It is also possible to modulate the contact force in the tangential plane \revc{to display} tactile effects if there is a relative \revc{displacement} between the finger and \revc{the} tactile surface. For example, due to the viscoelastic properties of finger pulp, the lateral movement of the tactile surface can induce tangential forces inside the finger pulp and thus \emph{lateral vibrations}. By modulating this lateral movement as a function of the finger displacement, it is even possible to render virtual textures on a tactile surface~\cite{wiertlewski:2011}. Indeed, by means of a phenomenon called \textit{causality inversion}, the forces produced by the lateral movement of the tactile surface can be matched to those produced by a real textured surface when a finger slides on it. 

Alternatively, researchers have devised ways to dynamically create \emph{friction modulation} between the tactile surface and the finger sliding on its surface. This can be, for example, achieved by using actuators that generate \emph{ultrasonic waves} on the tactile surface~\cite{Nara:2001, Biet:2007, Winfield:2007, ghenna:2017}. The vibration reduces the friction coefficient, a phenomenon called \textit{active lubrication}, because the finger is in intermittent contact with the plate. Using this approach, it is possible to simply turn the actuation on and off to generate relatively simple tactile effects in \revc{open loop} or to control the amplitude of vibrations in \revc{closed loop} to generate more complex tactile effects.

Instead of decreasing the friction via ultrasonic actuation, it is possible to increase it by using \emph{electrostatic} actuation~\cite{Grimnes83, Strong70, beebe1995polyimide, senseg:2008}. This involves applying a voltage to the conductive layer of a capacitive touchscreen to generate electrostatic attractive forces in the normal direction between its surface and a finger sliding on it, which leads to an increase in frictional forces applied to the finger opposite to the direction of movement. Although the magnitude of this electrostatic force in the normal direction is small relative to the normal load applied by the finger, it results in a perceivable frictional force in \revc{the} tangential direction when the finger slides on the touch screen. This frictional force can be modulated to generate different tactile effects by altering the amplitude, frequency and waveform of the voltage signal applied to the touchscreen.

In order to obtain a \emph{net tangential force} on the plane of \revc{the} tactile surface (i.e.\revc{,} vectorial summation of $F_t$ and $F_o$ in Fig.~\ref{fig:normal_vs_tangential}) even without any finger displacement, one can physically displace the tactile surface in the direction tangent to the finger movement on the same plane while modulating the friction using either ultrasonic or electrostatic actuation. This technique, named \textit{asymmetric friction}, can produce net forces with very low lateral displacements of the surface~\cite{Chubb2010}. Finally, by combining bending modes at high frequency of the surface, it is also possible to produce such a net force, by using \emph{elliptical movement} of the surface’s particles~\cite{ghenna:2017b}.

Figure \ref{fig-clas_Fig} presents the classification of the methods proposed in the literature to create surface haptics displays. The next section presents the type of actuators used to create tactile stimuli on touch surfaces, followed by \revc{an} in-depth discussion of the different stimulation methods given in this figure.

\subsection{\revc{Actuator Technologies}}

In most of the cases presented in Figure \ref{fig-clas_Fig}, the forces created at the contact area are produced through the medium that is constituted by the touch surface itself. Actuators can convert electrical power into mechanical power; they are placed in such a position that their action can control the contact forces.

There are many different types of actuators for generating tactile stimuli on touch surfaces. The \emph{electromagnetic actuators} generate a force which is proportional to a current. Among others, \emph{vibrations motors} have been used most widely for surface haptics applications. A vibration motor is a dc-motor that has a rotor with eccentric mass distribution (so also called ERM; eccentric rotating mass), which creates rotational movement to the housing and is in turn perceived as a vibration by the user. Vibration motors can generate relatively large vibrations and be made in diverse sizes and shapes. The frequency of vibration is controllable by changing the voltage. However, the amplitude (in displacement) is constant \cite{Ryu2010}, which restricts the diversity of vibration waveforms that can be rendered. \revc{Furthermore}, they generally have large actuation lags. Also popular are \emph{voice coil actuators}, which are similar to speakers used in audio systems. They can produce any waveforms with fast responses. Linear resonance actuators (LRAs) \revc{are} a special type of voice coil actuators designed in a very small size for mobile devices. While it has a \revc{faster response than an ERM}, its frequency bandwidth is generally very narrow.  In general, electromagnetic actuators are usually suitable for low-to-mid range actuation frequencies (with respect to human tactile perception), have low voltage requirements, and have high displacement/low force characteristics (see \cite{Choi2013} for \revc{a} detailed review).

\emph{Piezoelectric actuators} are also frequently used for surface haptics. They are more suitable for actuation of touch surfaces at high frequencies, and can generate high forces but at low displacement. They are also compact and can be easily mounted on a touch screen. However, they typically require very high voltage input, and are brittle and weak to external shock; major drawbacks preventing their commercial adoption.  

Those that have received attention more recently include \emph{electroactive polymers} -- like polyvinylidene fluoride --, which change their shape in volume when an electrical field is applied \revc{to} them. Voltage \revc{requirements are} often very high, but their softness allows them to be prepared on a large surface, in various sizes and forms. Finally, \emph{electrostatic actuators} produce a force which is proportional to the square of the voltage. The level of force is typically very low, but they can be used \revc{for} high frequency surface haptics applications. The main advantage here is that electrostatic forces can be generated on the \revc{finger's skin} directly, without taking into account the medium. 

Table \ref{tab:actuator} provides examples of the aforementioned actuator technologies for each method presented in Figure \ref{fig-clas_Fig}, while more details are given on their operating principles in the next sections.

\setlength{\extrarowheight}{1pt}
\begin{table*}[!ht]
 \caption{Examples of actuator technologies used for surface haptics}
\label{tab:actuator}
\begin{tabularx}{\textwidth}{@{}c*{4}{>{\centering}p{0.3cm}}*{6}{C}@{}}
Actuator technology     & \ro{Frequency} & \ro{Force output} & \ro{Displ. output} & \ro{Voltage req.} & Normal Vibration & Pulses & Lateral Vibration &  Friction Modulation & Driving Force & Asymetric Friction\\
\toprule
\textbf{Electromagnetic} & L & L & H & L & \cite{Yao:2010, Dhiab:2019} & \cite{Woo:2015, Bai:2011} & \cite{Aono:2014, Kess:2015} &  &  & \cite{Chubb2010, Mullenbach2012, imaizuni:2014}\\
\addlinespace
\textbf{Piezoelectric} & H & H & L & H & \cite{Hudin:2018, Dhiab:2019, Poupyrev:2002, Emgin19, Dai:2012b} & \cite{Hudin:2015} & \cite{Aono:2014, Wiertlewski:2013} & \cite{Biet:2007,ghenna:2017, Mullenbach2012, Hap2u, takasaki:2019} & \cite{dai:2012,ghenna:2017b} & \cite{Chubb2010, Mullenbach2012,xu:2019}\\
\addlinespace
\textbf{Electroactive Polymer} & L & L & H & H & \cite{Duong:2019, Frisson:2017} &  &  &  &  &\\
\addlinespace
\textbf{Electrostatic force} & H & L & L & H &  &  &  & \cite{senseg:2008, yamamoto2006electrostatic, tanvas} &  & \cite{xu:2019}\\
\bottomrule
\end{tabularx}
\end{table*}
\subsection{Force Modulation in the Normal Direction}

When a touch screen is actuated by a simple vibration motor as in mobile devices, a vibration wave propagating in the screen reaches the finger or hand in contact with its surface. Although this approach is sufficient for simple vibrotactile effects such as alerting the user about incoming call and receiving confirmation for button press, researchers have developed more sophisticated techniques to provide more interesting, useful, and salient vibrotactile effects, often using multiple actuators. Here we describe two such approaches: one is to generate illusory tactile sensations between the actual excitation positions and the other is to localize vibration to only \revc{a} small area on the touch surface. 

\subsubsection{Tactile Phantom Sensations on Touch Surfaces} \label{Phantom Sensations}

In general, the spatiotemporal characteristics of human tactile perception must be adequately reflected in designing surface haptics displays and associated interactions.  In spite of high human tactile acuity (spatial resolution of 1--2\,mm at the fingertip \cite{Goldstein02}), developing spatially-dense and distributed surface haptics displays is out of reach of current technology. One effective and popular way for generating such spatial tactile feedback is to exploit a perceptual illusion called \emph{phantom sensation} or \emph{funneling illusion}. A phantom sensation refers to an illusory tactile sensation that occurs midway between two or more distant tactile stimuli \cite{Allies1970}.  Phantom sensation is an effective method to improve the spatial resolution of a surface haptics display using only a few actuators.

According to the taxonomy of tactile phantom sensations \cite{Park2018}, those mediated through a rigid object such as \revc{a} touch surface are relevant to surface haptics. In this case, illusory tactile effects are perceived when the finger skin is in contact with the surface that is vibrated by multiple actuators simultaneously or in sequence. Such phantom sensations can be elicited on a line between two stimulation points (1D) or within a polygon surrounded by more stimulation points (2D). A phantom sensation can stay at one point (stationary) or move on the surface (moving). \revc{A moving} phantom sensation is similar in its notion to apparent tactile motion and sensory saltation \cite{Jones2006}. The perceived location of an illusory tactile sensation is controlled by adjusting either the amplitudes of the stimuli (amplitude inhibition) or their time gap (temporal inhibition) \cite{Allies1970}, or both.

Rendering stationary phantom sensations on a touch surface requires vibration damping mechanisms, such as those in \cite{RyuYangKang2009, YangRyuKang2009, Park2017, Park2019a}. This situation is similar to stimulating the skin directly, without vibrations generated by multiple actuators interfering with each other. It was demonstrated that stationary phantom sensations could be elicited on the surface of a mobile phone by controlling four vibration actuators with silicon-based dampers positioned at its corners \cite{Park2018}. The identification accuracy of the positions of the stationary phantom sensations was measured, and the distributions of the perceived positions were estimated. 

Without such dampers, only moving phantom sensations can be generated on a touch surface \cite{Park2018}, and most previous studies on surface haptics have focused on this type of stimulation. In this case, vibrations produced by multiple actuators are overlapped over the surface, and they are unable to provide spatially separate tactile cues required for eliciting stationary phantom sensations. Instead, we can vary the relative intensity or phase differences between the actuators over time, and such contrast effects lead to the perception of moving tactile sensations in spite of the fixed actuator positions. Kim et al.\revc{~\cite{KimSY2009}} attached two ERMs at the two long ends of a mobile phone and proposed a rendering algorithm for 1D moving phantom sensations based on temporal inhibition.
The authors then designed and fabricated a processor dedicated to the proposed rendering algorithm \cite{Kim2012}.
In contrast, Seo et al.\revc{~\cite{Seo2010}} used two LRAs and demonstrated 1D moving phantom sensations using amplitude inhibition.
They designed general synthesis functions using polynomials and examined the effects of their parameters on salient perceptual measures, such as sensation movement distance, velocity variation, intensity variation, and the confidence rating of moving sensation \cite{Seo2013}.
These results provide design guidelines of 1D moving phantom sensations with desired perceptual properties.
In \cite{KangKookChoEtAl2012}, \revc{Kang et al. used} piezoelectric actuators and demonstrated that \revc{the} amplitude profile has a significant effect on several perceptual qualities (accuracy, expression validity, and expression refinement).
They also carried out simulations and experiments for vibration propagation on a thin plate and showed that amplitude inhibition with frequency sweep led to the smoothest moving sensations \cite{Kang2012}.
Seo et al.\revc{~\cite{Seo2015_WHC}} recently extended their earlier work to a 2D case (called edge flows) by using four LRAs attached to the corners of a mobile phone to render moving phantom sensations following its edges.

\subsubsection{Localized Stimulation on Touch Surfaces}\label{localization}

In order to generate localized tactile effects on touch surfaces, the authors in~\cite{Dhiab:2019} \revc{altered} the geometrical parameters of the surface (especially the width to make it narrow) in order to confine the vibration at the location of the actuator. The resulting surface had the dimensions of $200\times 25\times 0.5$ mm$^3$. The solution proposed in this study is robust, but it is applicable to narrow plates and the actuator should be transparent so as not to obstruct the view. In \cite{Duong:2019}, the authors cover the entire surface with a relaxor ferroelectric polymer (RFP) transparent film, and produce stimulation based on the fretting phenomenon. They could create localized vibrations at a frequency of 500 Hz, with an amplitude of 1 $\mu m$ for a voltage amplitude of 200 V.

In~\cite{Emgin19}, the authors attached four piezoelectric patches to the edges of a touch screen to control \revc{out-of-plane} vibrations and display localized haptic feedback to the user. For this purpose, a vibration map of the touch surface was constructed in advance and then vibrotactile haptic feedback was displayed to the user using this map during real-time interaction with \revc{the} finger. The size of the touch surface was $743\times 447 \times 3$ mm$^3$.  In order to display localized haptic feedback, they divided the surface into 84 grid points and then maximized the vibration at the contact point, while minimizing it at the grid points around it. 

\revc{As an alternative} to the look-up table method suggested in~\cite{Emgin19}, vibration modes of a touch surface can be superimposed to create a vibration pattern that is detectable by \revc{the} fingertip. In \cite{Woo:2015}, 34 electromagnetic actuators, placed at the periphery of a glass plate with dimensions of $268 \times 170 \times 0.7$ mm$^3$, could control 34 vibration modes, and create a pulse on a square spot of $50 \times 50$ mm$^2$. To increase the contrast (the ratio of the vibration amplitude at the focus point to the amplitude anywhere else on the plate), the time reversal technique can be used. In \cite{Bai:2011}, for instance, the authors used 4 electromagnetic actuators to create a 3 ms duration pulse with an amplitude of 1 mm and a spot size of 20 mm in diameter on a glass plate with dimensions of $420 \times 420 \times 2$ mm$^3$. A higher contrast was obtained in \cite{Hudin:2015} with 32 piezoelectric actuators. The authors could create a pulse with an amplitude of 7$\mu m$ and a spot diameter of 5.2 mm on a plate with dimensions of $148 \times 210 \times 0.5$ mm$^3$.

\subsection{Force Modulation in \revc{the} Tangential Plane}
\subsubsection{Lateral Vibration}
In-plane vibrations of a touch surface can also produce tactile stimulation on the finger pad. Indeed, \cite{Wiertlewski:2013} shows that displacement of the skin by lateral vibrations can produce the same lateral forces as when interacting with a rough texture, and \revc{can elicit a similar sensation} of roughness for a virtual surface. 
The advantage of lateral vibrations is that actuators can be placed at the periphery of the touched surface, as in~\cite{Aono:2014}, so they do not have to be transparent. Electromagnetic ~\cite{Kess:2015} and piezoelectric~\cite{Aono:2014} actuators have been used for that purpose so far. 

\subsubsection{Friction Modulation}

Instead of producing a lateral force directly by using actuators that vibrate \revc{the} touch surface in \revc{the} tangential plane, it is possible to modulate the friction between the fingertip placed on a pad and \revc{the} touch surface using surface acoustic waves (SAW) as in \cite{nara:2000}. This solution is efficient only when the finger is moving on the surface, which is a limitation compared with lateral vibrations. However, the lateral force is obtained indirectly by friction modulation, which can be more easily achieved than the direct lateral force modulation. Alternatively, two more methods have been introduced in the literature to directly control the friction on a touch surface: by using electrostatic forces and ultrasonic vibrations. 

\textbf{Electrostatic Actuation.}
% Lead: CB
\revc{The electrical attraction between human skin and a charged surface is was first reported by Elisha Gray in a patent in 1875 ~\cite{gray1875improvement} and then Johnsen and Rahbek in 1923~\cite{Johnsen23}}. Approximately thirty years later, Mallinckrodt discovered by accident that dragging a dry finger over a conductive surface covered with a thin insulating layer and in the presence of the alternating voltage can increase friction during touch~\cite{Mallinckrodt53}. He explained this phenomenon based on the well-known principle of \revc{the} parallel plate capacitor. Later, Grimnes named this phenomenon electrovibration (electrically induced vibrations) and reported that surface roughness and dryness of finger skin could affect the perceived haptic effects~\cite{Grimnes83}. Afterwards, Strong and Troxel~\cite{Strong70} developed an electrotactile display consisting of an array of electrodes insulated with a thin layer of dielectric. Using friction induced by electrostatic attraction force, they generated texture sensations on the touch surface. Beebe et al.~\cite{beebe1995polyimide} developed a polyimide-on-silicon electrostatic fingertip tactile display using lithographic microfabrication. They were able to generate tactile sensations on this thin and durable display using 200--600 V voltage pulses and reported the perception at the fingertip as sticky. Later, Tang and Beebe~\cite{tang1998microfabricated} performed experiments for detection threshold, line separation and pattern recognition with visually impaired subjects. The subjects successfully  differentiated simple tactile patterns by tactile exploration. 

\begin{figure}[ht]
\centering
\includegraphics[width=1.0\columnwidth]{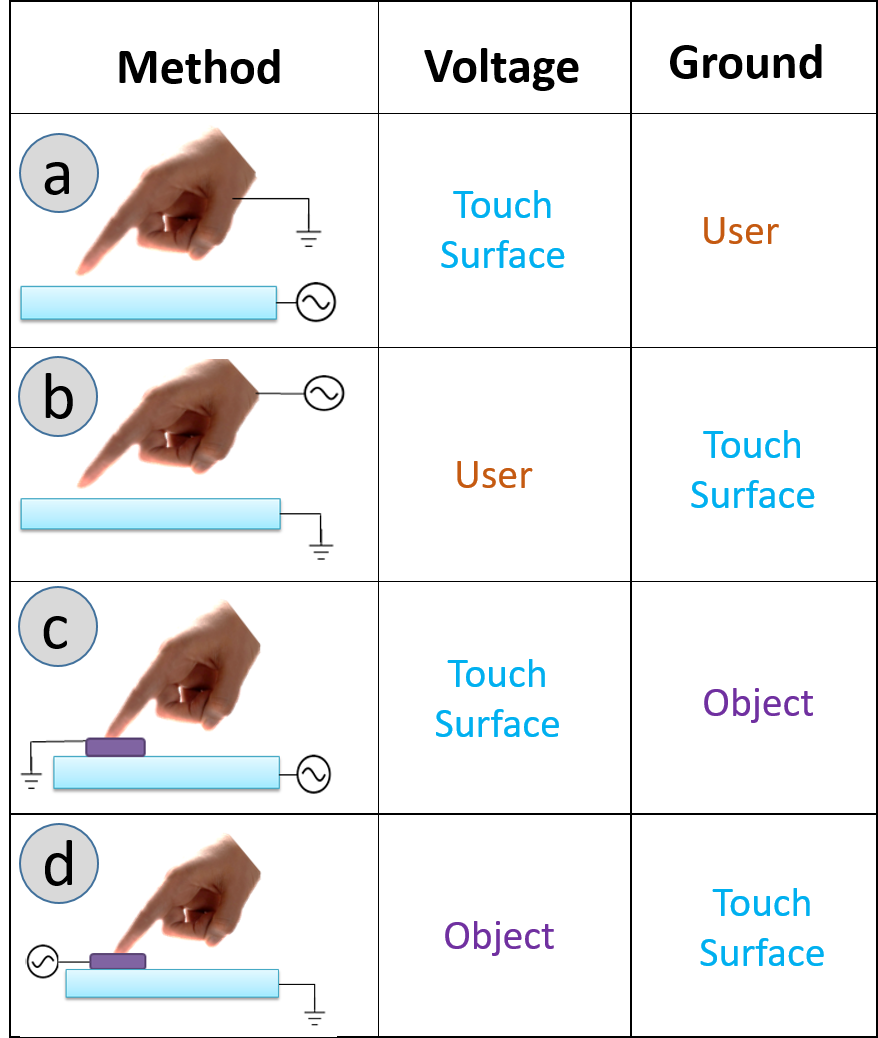}
\caption{Different methods of grounding for electrostatic actuation}
\label{fig:different_grounding_electrostatic}
\end{figure}

More recently, Bau et al.~\cite{Bau10} and Linjama et al.~\cite{linjama2009sense} showed that electrovibration can be delivered through a commercial capacitive touch screen, which demonstrates the viability of this technology on \revc{large surfaces}. The surface capacitive touchscreen used in those studies consists of a glass substrate, coated with Indium Tin Oxide (ITO) as a conductive layer and Silicon Dioxide ($SiO_2$) as an insulating layer. When an alternating voltage is applied to its conductive layer (see Fig.~\ref{fig:different_grounding_electrostatic}a), as the human body is also an electrical conductor, touching the surface of the screen results in an attraction force between finger and surface of touchscreen due to the induction of charges with opposite signs on the insulating layer of the glass substrate and the finger. In fact, this is only one of the ways of generating electrostatic attractive force between \revc{a} human fingerpad and a touch surface. Bau and Poupyrev~\cite{Bau12} proposed REVEL, in which the voltage was applied to the user's finger rather than the conductive layer of the touschscreen (Fig.~\ref{fig:different_grounding_electrostatic}b). Hence, REVEL is based on the principle of, as the authors call it, \textit{reverse electrovibration} and enables to change the tactile feeling of real touch surfaces by augmenting them with virtual tactile textures. 

It is also possible to generate electrostatic attractive forces between \revc{a} finger and an object sliding on a touch surface (Fig.~\ref{fig:different_grounding_electrostatic}c). For example, Yamamoto et al.~\cite{yamamoto2006electrostatic} developed an electrostatic tactile display that consists of a conductive pad sliding on a touch surface with embedded conductive electrodes. The sliding pad was not explicitly grounded, but a pair of positive and negative alternating voltage signals were applied to the electrodes to make the voltage balance around 0 V on the surface of \revc{the} pad. The display is incorporated into a tactile telepresentation system to enable exploration of remote surface textures with real-time tactile feedback to the user. Later, Nakamura and Yamamoto~\cite{Nakamura16} developed a multiple user system by applying voltage to the sliding pad and grounding the touch surface (Fig.~\ref{fig:different_grounding_electrostatic}d). This technology could in fact open doors to some exciting multi-user applications in gaming, education, \revc{and} data and information visualization on large scale touch surfaces. For example, a multi-user virtual hockey game was implemented as a demonstration of the technology in~\cite{Nakamura16}. 

Due to variations in human and environmental impedances, the voltage-induced stimulation may cause tactile feedback with nonuniform intensity on touch surface. Alternatively, a current feedback method was proposed in~\cite{kim2015method} and the results of the user study showed that it could provide significantly more \revc{uniformly} perceived intensity of electrovibration as compared with the conventional voltage control method.
% This table is put here to appear on the same page as Ultrasonic Waves section
\setlength{\extrarowheight}{1pt}
\begin{table*}[!ht]
 \caption{Examples of haptic surfaces based on ultrasonic actuation.}
\label{tab:devices}
\centering
\begin{tabularx}{0.9\textwidth}{@{}c c c *{4}{C}@{}}
Year     & Device & Size & Material & Voltage (peak value)& Frequency & Vibration amplitude \\
\toprule
2007 & StimTac~\cite{Biet:2007} & $83 \times 49$ mm$^2$ & Copper & 15 V & 34.77 kHz & 2.3
$\mu$m \\
2007 & TPad~\cite{Winfield:2007}   & \O25 mm & Glass & 70 V & 30 kHz & \\
2010 & LATPad~\cite{Marchuk:2010} & $75 \times 75 $ mm$^2$ & Glass & 60 V & $\sim 27$ kHz &  \\
2012 &   \cite{giraud:2012}     & $93 \times 65$ mm$^2$ & Glass & 150 V & 31.2 kHz & 1.5
$\mu$m \\
2013 &    \cite{casset:2013}    & $60 \times 40$ mm$^2$ & Si Wafer & 4 V   & 31.4 kHz  & 1.1
$\mu$m\\
2013 & TPadFire~\cite{Mullenbach:2013} & $165\times130$ mm$^2$ & Glass & 100 V & 33.5 kHz & 1.2
$\mu$m  \\
2014 & Haptic Sensory Tablet~\cite{fujitsu} & $270 \times 177$mm$^2$& Glass & & & \\
2015 &   \cite{Yang:2015}     & $198 \times 138$ mm $^2$ & Alu. & 20 V & 52.4 kHz & 1.25
$\mu$m\\
2016 & eVITA~\cite{vezzoli:2016}  & $154 \times 81$ mm$^2$ & Glass & 25 V & 60 kHz & 1
$\mu$m\\
2018 &   \cite{giraud:2018}     & $110 \times 19$ mm$^2$ & Glass & 45 V & 225 kHz & 0.125
$\mu$m\\
\bottomrule
\end{tabularx}
\end{table*}

While the technology for generating tactile feedback on a touch screen via electrovibration is already in place and straightforward to implement, our knowledge on the underlying contact mechanics and tactile perception are highly limited~\cite{sirin2019fingerpad}. This is not surprising since the contact interactions between human skin and a counter surface is already highly complex to investigate even without electrovibration~\cite{tomlinson2009understanding},~\cite{adams2013finger},~\cite{van2015review},~\cite{sahli2018evolution}. In the case of electrovibration in particular, the exact mechanism leading to an increase in tangential frictional forces is still not known completely. 

To estimate the electrostatic forces between finger pad and touch screen, models based on parallel-plate capacitor theory have been proposed (see the summary in~\cite{vodlak2016multi}). These models assume a constant air gap between two surfaces and ignore the rounded shape of the finger pad and the asperities on its surface. On the other hand, to estimate the frictional forces in tangential direction during sliding, the Coulomb model of friction utilizing a constant coefficient of dynamic friction between finger pad and touch screen has been used so far. The increase in tangential force has been explained by simply adding the force due to electrostatic attraction to the normal force applied by the finger, $F_t=\mu(F_n+F_e)$, where $\mu$ is the friction coefficient and $F_e$ represents the electrostatic force. 

However, friction of human skin against smooth surfaces is governed by \revc{the} adhesion model of friction~\cite{bowden2001friction,Adams07}, which depends on interfacial shear strength and real area of contact as $F_t=\tau A_{real}$. It is highly difficult to measure or estimate the real area of contact, $A_{real}$, which varies nonlinearly with the normal force. In earlier studies, the real contact area has been taken as the contact area of finger ridges, $A_{ridge}$, which is smaller than the apparent contact area, $A_{apparent}$, estimated by the fingerprint images using the boundaries of \revc{the} finger pad in contact with \revc{the} surface~\cite{bochereau2017characterizing}. On the other hand, if the Persson’s multi-scale contact theory is considered (see the review of contact theories in ~\cite{persson2006contact}), the real area of contact ($A_{real}$) is, in fact, much smaller than the contact area of finger ridges ($A_{ridge}$) since only the finger asperities at finer resolution make contacts with the surface. Recently, Persson extended his theory to electroadhesion and derived a mathematical relation for electroadhesive pressure between two conductive surfaces with random roughness in contact for voltage applied to one of the conducting surfaces~\cite{persson2018dependency}. He assumed the small slope approximation for the roughness profiles and solved the Laplace equation for electrostatic potential to obtain the pressure due to electroadhesive force for the applied voltage. He used the normal pressure to estimate the real contact area and then the tangential friction force using the adhesion friction model given above. 

The application of this approach to the interactions between human finger and touch screen under electroadhesion has been investigated by Ayyildiz et al.~\cite{ayyildiz2018contact} and Sirin et al.~\cite{sirin2019electroadhesion} and verified by the experimental data collected via a custom-made tribometer. These studies show that electrostatic attraction force increases the \emph{real contact area} between fingerpad and  touchscreen, leading to an increase in sliding friction force. The authors argue that the artificially generated electrostatic force results in a decrease in interfacial gap and hence the finger asperities make more contacts with the surface of the touch screen at the microscopic scale, leading to an increase in real contact area of \revc{the} finger. Supporting this claim, Shultz et al. ~\cite{shultz2018electrical} showed that the electrical impedance of the interfacial gap is significantly lower for the stationary finger compared to that of the sliding finger under electroadhesion. In contrast to the argument on increase in real contact area, which is highly difficult to measure and validate experimentally, the decrease in apparent contact area under electrovibration has been already demonstrated experimentally. Sirin et al.~\cite{sirin2019fingerpad} conducted an experimental study to investigate the contact evolution between the human finger and a touch screen under electrovibration using a robotic set-up and an imaging system. The results show that the effect of electrovibration is only present during full slip but not before slip. Hence, the coefficient of friction increases under electrovibration as expected during full slip, but the apparent contact area is significantly smaller during full slip when compared to that of \revc{the condition without electrovibration}. It was suggested that the main cause of the increase in friction during full slip is due to an increase in the real contact area and the reduction in apparent area is due to stiffening of the finger skin in the tangential direction.

\textbf{Ultrasonic Actuation.}
When a finger slides over a high frequency vibrating plate, the friction that \revc{this motion produces} decreases as the vibration amplitude increases. This phenomenon, called \emph{active lubrication}, has been first introduced by \cite{Watanabe95} to develop a tactile stimulator. So far, two different mechanisms have been suggested to explain the cause of friction reduction in ultrasonic tactile displays. \cite{Biet:2007} proposed that the friction reduction is due to the formation of a squeeze film of air between \revc{the} finger and the surface. Alternatively, it has been suggested that, when a surface vibrates at an ultrasonic frequency, an intermittent mechanical contact develops between finger and the surface such that the finger bounces on the surface while sliding~\cite{Vezzoli:2017}. A recent study conducted by Wiertlewski et al. \cite{Wiertlewski16} using a stroboscope revealed that both mechanisms indeed contribute to the friction reduction. Hence, it appears that the fingertip bounces on a cushion of squeezed film of air. On the other hand, another recent study shows that keeping the vibration acceleration constant rather than the vibration amplitude leads to more uniform air-gap thickness and hence allows for constant reduction in friction~\cite{giraud:2018}.

To generate ultrasonic vibration on a touch surface, piezoelectric film actuators are typically glued on the surface and actuated by sinusoidal voltage signals at a resonance frequency above 20 kHz~\cite{Marchuk:2010}. To efficiently produce this vibration, a stationary bending mode is energized, and \revc{a} specific design procedure is carried out to optimize the power efficiency \cite{giraud:2010, Wiertlewski15} and to free the touch surface from non-transparent actuators. In \cite{giraud:2012}, for example, 400 mW of power was required for a voltage amplitude of 150 Vpp to produce a vibration amplitude of 1.5 $\mu$m at 31 kHz on a $93 \times 65 \times 0.9$ mm$^3$ glass plate. These advances make it possible to create tablets that are able to provide users with tactile feedback via friction modulation~\cite{ Hap2u, Mullenbach:2013}. 

Voltage \revc{requirements are} typically higher for vibrating a glass plate than a metal substrate of the same size and shape since the glass plate vibrates with more internal damping. To reduce the voltage requirement, it is possible to reduce the piezoelectric material thickness or deposit the piezoelectric active material directly on the %glass
substrate. In \cite{casset:2013}, for example, the authors obtained a vibration amplitude of 1.1
$\mu$m for a voltage amplitude of 8 Vpp on a silicon %glass
substrate of size $60 \times 40 $ mm$^2$. 

The combination of two vibration modes is also possible at ultrasonic frequency. This can lead to dual-touch tactile stimulator, as in \cite{ghenna:2017}, where two vibration modes are controlled in amplitude and phase in order to differentiate the vibration at two points of the plate.

 Ultrasonic actuation can provide a higher variation in friction as compared to the electrostatic actuation, while the rendering bandwidth of ultrasonic devices is limited due to its resonating nature~\cite{Vardar17}. Obviously, there exist a trade-off between the energy efficiency of a device and its bandwidth. To cope for this issue, it is possible to increase the operating frequency, then the necessary bandwidth of the device will be small relative to its resonance frequency, and hence the quality factor of the system will be high. In ~\cite{takasaki:2019}, for example, the authors obtained a quality factor of 775 on a plate made of piezoelectric material resonating at 881 kHz. It is also possible to introduce a feed-forward  control as in~\cite{wiert:2014} to compensate for the undesired signal attenuation behaviour of the plate. Furthermore, closed-loop controls can be used to dynamically adapt the level of voltage amplitude in order to compensate for the damping effect of finger on the plate, yet tracking the resonance frequency of the system to enable efficient contact interactions~\cite{vezzoli:2016, messaoud:2016b}.

\subsection{Net Tangential Force}
\label{section:net-tangential-force}

It is possible to create shear forces on the bare finger that can be controlled independently of finger speed and direction. In \cite{Chubb2010}, an \emph{asymmetric friction} force was created by a device involving a touch surface that was able to produce active lubrication by ultrasonic actuation via piezo patches and a electromagnetic motor oscillating this surface at 100 Hz in \revc{the} tangential direction. By synchronizing the friction reduction on the oscillation, forces up to 100 mN could be displayed to the user in this system. 
In a similar approach, \cite{Mullenbach:2016} uses electrovibration with oscillations at 1000 Hz to produce lateral forces up to 0.45 N. To obtain controllable forces on a touch surface in both directions ($F_t$ and $F_o$ in Fig.~\ref{fig:normal_vs_tangential}), \cite{Mullenbach2012} developed a large area variable friction device, mounted on an impedance controlled planar mechanism. In this way, static and dynamic frictional forces can be controlled as well as the transition between the two regimes. To avoid operating at high frequencies, \cite{imaizuni:2014} designed a vibrating surface that follows a saw-tooth like displacement profile with a low frequency of 5 Hz. In this system, the participants could feel more friction when sliding their finger in one direction compared to the other. This solution needs a high amplitude and low frequency movement of the touched surface, which is a limitation of this principle. To cope with this issue, authors in \cite{xu:2019} developed a surface haptics display that can simultaneously modulate friction via electrostatic actuation and produce in-plane ultrasonic oscillations at 30 kHz. This system could generate active lateral forces up to 0.4 N.
The direction and magnitude of the lateral force can be adjusted by varying the phase between the input signals of in-plane oscillation and the electroadhesion.

It is also possible to generate a \emph{driving force} directly by using two different modal behaviors of a touch surface, promoted at the same time by exciting them at an intermediate frequency between the corresponding two resonance frequencies. For instance, \cite{dai:2012} combined an out-of-plate vibration mode of a touch surface with an in-plane lateral one, both having the same resonant frequency (22.3 kHz), to obtain a tilted straight-line motion of the touch surface. The force that develops on the finger is controlled by modulating the relative phase of the two resonances. In this system, lateral forces up to 0.05~N could be exerted on a fixed bare finger. An elliptical motion of the particles, which occurs when a travelling wave is propagating over the plate, can also produce a driving force~\cite{ghenna:2017b}, provided that the friction force can be compensated. For instance, a driving force up to 0.05 N could be generated by using a travelling wave with a frequency of 28,4 kHz and an amplitude of 1.6
$\mu$m. This principle has been validated on a beam, but seems difficult to translate to a glass plate.
%However, it is possible to implement electrostatic and ultrasonic actuation techniques together on the same surface to achieve even larger variation in friction coefficient [8]~\cite{MISSING}.
\section{Human Haptics}

This section covers the studies on surface haptics investigating human tactile perception and contact mechanics between the fingerpad and touch surfaces displaying tactile feedback. As in any other tactile stimuli applied to the finger, it is encoded by the mechanoreceptors first. The encoded signals then travel across a series of neural structures and reach the brain to form the percept. 

Due to the limitations in machine haptics today, haptic interactions of a finger with a touch surface \revc{are} mostly limited to the \emph{tactile} channel. In surface haptics, the surface is generally immovable during interaction. It is extremely challenging to design and implement a surface that makes clearly perceptible displacement, either normal or tangential, while satisfying other requirements such as protection from dust or water. This implies that it is safe to ignore the role of \emph{kinesthetic} perception in most cases and focus on tactile perception. In addition, pressure perception is irrelevant since it requires substantial deformation of the skin. Therefore, vibrotactile and frictional stimuli have the most prevalent effects in surface haptics. In this regard, we review the literature based on these two kinds of stimuli in the upcoming \revc{subsections}. 

\subsection{Vibrotactile Stimuli}

To achieve a target perceptual effect, we generate a tactile stimulus using an actuator and deliver it \revc{to} the user's finger when the finger taps, presses, or slides on the surface. Here the user's finger is already in contact with the surface. The natural mechanical response from the surface always occurs, and it is mixed with the synthetic tactile stimulus. This corresponds to the situation of \emph{haptic augmented reality} \cite{Jeon2009} though it has  been rarely acknowledged in the literature. Therefore, to elicit desired tactile effects, we need salient artificial tactile stimuli that are distinct from the natural tactile stimuli provided by the surface. This is one of the reasons why vibrotactile stimuli have been preferred in the majority of surface haptics applications so far. 

The modern theories on vibrotactile perception are based on the four channel theory, which states that glabrous (non-hairy) human skin contains four types of mechanoreceptors (sensory end-organs that detect skin deformation)~\cite{Bolanowski88, Gescheider2008}. If stimulated, each mechanoreceptor fires an electrical pulse, which is transmitted to the celebral cortex through \revc{a} nerve. Thus, humans have four kinds of mechanoreceptive channel, and they are classified by sensory adaptation (slow or rapid) and receptive field (small or large). Among the four, two types of fast-adapting (FA) mechanoreceptive channel are in charge of vibrotactile perception. These mechanoreceptors mainly respond to the changes in a stimulus, not much to the steady-state value, due to quick adaptation to the stimulus. This adaptation behavior makes the \revc{FA} channels suitable for perceiving the time-varying properties of the stimulus.    

The PC (Pacinian) channel, also called FA (fast adapting) II or P channel, has an end organ called Pacinian corpuscle. The Pacinian corpuscle responds to vibrating mechanical stimuli in a wide frequency band (10--1000 Hz). Its sensitivity depends on many factors. For example, the PC channel has the highest sensitivity at a frequency between 200 Hz and 300 Hz. The receptive field of Pacinian corpuscle is relatively large, and so its spatial acuity is quite low. People generally describe the subjective impression of vibration mediated by the PC channel as a smooth vibration. 

The PC channel is capable of energy summation both temporally or spatially. In the temporal domain, a longer stimulus has a lower absolute threshold and so more chance to be perceived than a shorter stimulus of the same intensity. Likewise, a longer stimulus is perceived to be stronger than a shorter stimulus. The same applies to the contact area of a stimulus on the skin, which is called the spatial summation property. Therefore, stimulus duration and contact area are the key parameters that determine various perceptual characteristics of the PC channel. 

\revc{The other fast-adapting channel is the FA I or NP (Non-Pacinian) I channel. This channel has a Meissner corpuscle as the receptor}, which is activated by relatively low-frequency (3--100 Hz) vibrations. The Meissner corpuscle has a receptive field smaller than the Pacinian corpuscle, and the \revc{FA I} channel has better spatial resolution than the PC channel. The \revc{FA I} channel also has quite different properties from the PC channel. For example, its detection thresholds are not critically dependent on vibration frequency. The \revc{FA I} channel is not subject to temporal or spatial summation. The low-frequency vibrations mediated by the \revc{FA I} channel have a clear fluttering sensation. 

The perception of vibrotactile stimuli depends on a large number of factors, such as frequency, amplitude, waveform, body site in contact, duration, contact area, and age. There exist many comprehensive and in-depth reviews on vibrotactile perception \cite{Choi2013, Goldstein02,Jones2006,Gescheider2008,Jones2008,Johansson2009}, and we recommend them for interested readers. 

\revn{A few notes are to be made. First, the Pacinian corpuscles do not discriminate the stimulation direction of vibration \cite{Brisben99}. Thus, the perceptual properties between vibrations in the directions normal and tangential to the surface are very similar.  Second, there are fundamental differences between passive touch and active touch. What is described so far is the basis of passive perception of vibrotactile stimuli, where no motor commands are involved. Passive touch is the case when a hand holds a mobile phone or a finger maintains contact with a touchscreen. When hands or fingers move, e.g., for friction perception, underlying perceptual mechanisms and processes are much more complex. }

%\hl{SEUNGMOON: Mechanotransduction of Vibrotactile Stimulation in finger?}
%\begin{itemize}
%\item Impedance in normal and tangential direction
%\item Propagation of vibration on the finger (Yon has done some work)
%\item Effect of finger skin material properties on tactile feeling of vibrotactile stimulus in normal and letral direction
%\item Neuroscience/Mechanotransduction aspects?
%\end{itemize}
%\hl{SEUNGMOON: vibrotactile (NORMAL and LATERAL DIRECTION) studies on surface haptics displays will be reported here}\\
\subsection{Frictional  Stimuli}

Compared to vibrotactile stimuli, tactile perception of frictional stimuli has been studied much less, and little is known about the sensory and neuronal processes behind it. During friction perception, a relative sliding motion between the finger and the surface occurs. There both the normal and tangential forces play a role, and they generate compound strains (both normal and shear) on the fingerpad.

There are several parameters affecting the tactile perception of friction. Obviously, the roughness and material properties of the surface \revc{with} which the fingerpad is in contact are highly important. The earlier studies in \revc{the} tribology literature show that friction between skin and a smooth glass surface such as a touch screen is mainly governed by \emph{adhesion}, which is effective at lower normal contact forces~\cite{Adams07}. Increasing normal force increases tangential force, while the coefficient of friction decreases and eventually \revc{reaches} a steady state value~\cite{Derler09}. However, it is important to note here that the friction between a smooth surface and \revc{a} fingerpad varies with many other finger-related parameters such as moisture, velocity, mechanical properties, and fingerprints as well as age and gender (see the reviews in ~\cite{adams2013finger, derler2012tribology}). For example, in interactions with smooth surfaces, accumulated finger moisture at interfacial gap results in softening of the fingerpad, known as plasticization~\cite{adams2013finger, Adams07}, leading to an increase in coefficient of friction. 

In the next two \revc{subsections}, we cover the tactile perception of stimuli displayed by electrostatic and ultrasonic actuation since those are the two prominent approaches for friction modulation on touch surfaces.    

\textbf{Perception of Electrovibration Stimuli.}
The studies on tactile perception of frictional stimuli on touch surfaces have mainly focused on the estimation of perceptual thresholds for periodic stimuli and its roughness perception so far. As discussed earlier, electrostatic and ultrasonic actuation are typically used to modulate friction on a touch surface. Bau et al. measured the sensory thresholds of electrovibration using sinusoidal inputs applied at different frequencies~\cite{Bau10}. They showed that the change in threshold voltage as a function of frequency followed a U-shaped curve similar to the one observed in vibrotactile studies. Later, Wijekoon et al.~\cite{wijekoon2012electrostatic} investigated the perceived intensity of modulated friction generated by electrovibration. Their experimental results showed that the perceived intensity was logarithmically proportional to the amplitude of the applied voltage signal. Vardar et al.~\cite{Vardar17, Vardar16} investigated the effect of input voltage waveform on \revc{the} tactile perception of electrovibration. They showed that humans were more sensitive to tactile stimuli generated by square wave voltage than sinusoidal one at frequencies below 60 Hz. In order to interpret the outcome of these experiments, the force and acceleration data collected from the subjects were analyzed in the frequency domain by taking into account the human vibrotactile sensitivity curve. This detailed analysis showed that \revc{the} Pacinian channel was the primary psychophysical channel responsible for the detection of the electrovibration stimuli, which is most effective \revc{for} tactile stimuli at high frequencies around 250 Hz. Hence, the stronger tactile sensation caused by a low-frequency square wave was due to its high- frequency components stimulating the Pacinian channel.

Vardar et al.~\cite{Vardar18} investigated the interference of multiple tactile stimuli (\emph{tactile masking}) under electrovibration. They conducted psychophysical experiments and showed that masking effectiveness was larger with pedestal masking than simultaneous one. They also showed that sharpness perception of virtual edges displayed on touch screens depends on the \emph{haptic contrast} between background and foreground stimuli, similar to the way it has been observed in vision studies, which varies as a function of masking amplitude and activation levels of frequency-dependent psychophysical channels. 

\revn{Ryo et al. \cite{ryu2018mechanical} attached piezoactuators to a capacitive touch screen and investigated the effect of \emph {in-site} vibrotactile masking on tactile perception of electrovibration. They showed that the absolute threshold of electrovibration increases in the form of a
ramp function as the intensity of the masking stimulus (mechanical vibration) increased.
The masking effect was more prominent when the frequency of both the target and the masking
stimulus was the same.} Jamalzadeh et al. ~\cite{jamalzadeh2019effect} investigated whether it is possible to change the detection threshold of electrovibration at fingertip of \revc{the} index finger via \emph{remote masking}, i.e. by applying a (mechanical) vibrotactile stimulus on the proximal phalanx of the same finger. The masking stimuli were generated by a voice coil (Haptuator). The results of their experimental study with 8 participants showed that vibrotactile masking stimuli generated sub-threshold vibrations around \revc{the} fingertip and, hence, did not mechanically interfere with the electrovibration stimulus. However, there was a clear psychophysical masking effect due to central neural processes. \revc{The electrovibration} absolute threshold increased approximately 0.19 dB \revc{per} dB increase in the masking level. 

The number of studies on roughness perception of periodic stimuli rendered on a touch surface by electrovibration is only a few and the underlying perceptual mechanisms have not been established yet. Vardar et al.~\cite{vardar2017roughness} investigated the roughness perception of periodic gratings displayed by electrovibration. They compared four wave-forms\revc{:} sine, square, triangular and saw-toothed waves with spatial period varying from 0.6 to 8 mm. The width of periodic high friction regimes (analogous to ridge width) was taken as 0.5 mm, while that of the low friction ones (analogous to groove width) was varied. The results showed that \revc{the} square \revc{waveform} was perceived as the roughest, while there was no significant difference between the other three waveforms. The perceived roughness decreased with increasing spatial period in general, though a modest U-shaped trend was observed with a peak value around 2~mm. They also reported that the roughness perception of periodic gratings correlates best with the rate of change in tangential force, similar to real gratings~\cite{smith2002role}. Isleyen et al.~\cite{Isleyen19} compared the roughness perception of real gratings made of plexiglass with virtual gratings displayed by electrovibration through a touch screen. The results of their experimental study clearly showed that the roughness perception of real and virtual gratings are different. They argued that this difference can be explained by the amount of fingerpad penetration into the gratings. For real gratings, penetration increased the tangential forces acting on the finger, whereas for virtual ones where skin penetration is absent, the tangential forces decreased with spatial period. Supporting this claim, they also found that increasing normal force increases the perceived roughness of real gratings while it causes an opposite effect for the virtual gratings. \revn{Ito et al. \cite{ito2019tactile} investigated the effect of combining vibrotactile and electrostatic stimuli in different ratios on tactile roughness perception of virtual sinusoidal gratings. Their experimental results showed that a vibrotactile stimulus with a slight variable-friction stimulus is perceptually more effective for displaying gratings with surface wavelengths greater than or equal to 1.0 mm.}

\revn{Recently, Ozdamar et al. \cite{ozdamar2020step} investigated the tactile perception of a step change in friction and the underlying contact mechanics using a custom-made set-up including a high speed and frame-rate camera for finger imaging. The experimental results showed that the participants perceived rising friction stronger than falling friction, and both the normal force
and sliding velocity significantly influenced their perception. These results were supported by the tribological measurements on the relative change in
friction, the apparent contact area of the finger, and the elastic strain acting on the finger in the sliding direction.}

The study by Mun et al. \cite{Mun2019} was concerned with the perceptual structure of haptic textures rendered by an electrovibration display. Using 32 textures expressed using regular tessellations of polygons, they obtained a three-dimensional perceptual space that visualized the pairwise perceptual dissimilarities between the textures and their resulting structure. In addition, three perceptual dimensions most adequate to
describe the perceptual attributes of the electrovibration textures were found to be rough-smooth,
dense-sparse, and bumpy-even.

\textbf{Perception of Ultrasonic Stimuli.}
Compared to electrovibration, generating controlled stimuli using ultrasonic actuation to conduct psychophysical experiments is more difficult. Samur et al.~\cite{samur2009psychophysical} conducted discrimination experiments to evaluate the minimum detectable difference in friction using the tactile pattern display (TPad), actuated at an ultrasonic resonance frequency. The subjects were presented with two stimuli in sequential order and asked to identify the stimulus with higher friction. An average JND of 18\% was reported for the friction difference. This study shows the ability of humans to discriminate two surfaces based on friction, but how humans perceive a change in friction cannot be ascertained. 

In this regard, Messaoud et al.~\cite{messaoud2016relation} have evaluated the subjects' performance in detecting a change in friction. Their results showed that the detection rate improves at lower inherent friction between finger and surface, as well as lower finger velocity of 5 mm/s compared to 20 mm/s. They also found that the detection rate is best correlated to friction contrast and a contrast greater than 0.19 is always detectable. It was shown in Saleem et al. ~\cite{saleem2017tactile} that rate of change in tangential force is best correlated with the detection rate of friction change. Later, Saleem et al. ~\cite{Saleem18} also showed that a step increase in friction by ultrasonic actuation casts a stronger perceptual effect as compared to a step fall in friction, supporting the recent results of \cite{ozdamar2020step} obtained by electrostatic actuation. Gueorguiev et al.~\cite{gueorguiev2017feeling} investigated the tactile perception of square pulses displayed by ultrasonic friction modulation and observed that subjects could differentiate between two pulses if the duration and transition time are extended by 2.4 and 2.1~ms, respectively. In another experiment, they found that if the duration between two pulses was shorter than 50 ms, subjects perceived them as a single pulse. 

In~\cite{gueorguiev2017tactile}, the authors evaluated the threshold to detect two friction reductions of 100 ms duration, rendered 3 mm apart, using three different ultrasonically actuated surfaces made of aluminum, polypropylene and polyurethane. The rise time of vibration amplitude was controlled at 1.5 ms. The surfaces were passively scanned under the finger at 20 mm/s, while normal force was maintained at 0.7 N. They conducted threshold experiments and measured the vibration thresholds at 75\% JND as 0.17, 0.23, 0.27 $\mu$m for aluminum, polypropylene and polyurethane, respectively. Furthermore, the detection rate was found to correlate well with the ratio of reduction in tangential force to pre-stimulation tangential force. 

The discrimination of periodic gratings, rendered on a touch surface by ultrasonic actuation was investigated in~\cite{biet2008discrimination}. The surface was vibrated at a frequency of 30 kHz with an amplitude of 1.1 $\mu$m. Four standard and eight comparison gratings in square waveform were rendered. The width ratio of high to low friction regimes was always kept constant. The finger velocity was not controlled. The results showed that the difference threshold (JND) increases with spatial period (0.2, 0.32, 0.47 and 0.8 mm for spatial period of 2.5, 3.5, 5 and 10 mm, respectively). Unlike real gratings [44], \revc{the} Weber fraction remained almost constant (varied between 8 and 10 \%). 

More recently, Saleem et al~\cite{saleemtactile} investigated our ability to discriminate two consecutive changes in friction (called edges), followed by discrimination and roughness perception of multiple edges (called periodic gratings). The results showed that discrimination of two consecutive edges was significantly influenced by edge sequence: a step fall in friction (FF) followed by a step rise in friction (RF) was discriminated more easily than the reverse order. On the other hand, periodic gratings generated by displaying consecutive edge sequence of FF followed by RF were perceived with the same acuity as compared to vice versa. They also found that a relative difference of 14\% in spatial period was required to discriminate two periodic gratings independent of the edge sequence. Moreover, the roughness perception of periodic gratings decreased with increasing spatial period for the range that they have investigated (spatial period $<$ 2 mm), supporting the results observed by Vardar et al. ~\cite{vardar2017roughness} and Isleyen et al. ~\cite{Isleyen19} for friction modulation by electrovibration.

%%%%%%%%%%%%%%%%%%%%%%%%%%%%%%%%%%%%%%%%%%%%%%%%%%%
\section{Human-Machine Haptics}

Touch-enabled devices such as smartphones and tablets are ubiquitous and used on a daily basis. The devices that are commercially available, however, provide visual and auditory feedback but almost no tactile feedback, even though it is known that tactile feedback improves task performance and realism when interacting with digital data. Moreover, tactile sensation is a significant factor in the preference and positive attitude towards consumer products due to their surface texture. Therefore, a great deal of research is being carried out in recent years to develop and study the techniques that can render tactile information on touch-enabled devices.

This section mainly covers the tactile rendering algorithms for displaying virtual textures and shapes on touch surfaces, the design of user interfaces and experiences with surfaces haptics, and some current applications of surface haptics. A summary \revc{of how the technologies} of surface haptics, as described in Figure \ref{fig-clas_Fig}, \revc{are} used to implement these techniques is given in Table \ref{tab:HM}. 

%%%%%%%%%%%%%%%%%%%%%%%%%%%%%%%%%%%%%%%%%%%%%%%%%%%
\subsection{Tactile Rendering}

The rendering of virtual textures, shapes and surface features is important for many applications of surface haptics. Virtual textures can for example be used to simulate material properties or provide informative feedback. Virtual shapes and surface features can similarly be used to augment spatial elements such as buttons or graphical content, indicating location, function, state or other properties. Each type of tactile rendering will be discussed in turn.

%%%
\subsubsection{Rendering of Virtual Textures}

Friction modulation is particularly effective at texture rendering due to its frequency range and inherently passive nature. Virtual textures produced by friction modulation are felt only while a finger brushes against them, thereby increasing their realism. Importantly, this passivity is a fundamental property of friction modulation and is independent of the responsiveness of the display's actuation.

Virtual textures can be synthesized by actuating a friction display with a time-varying periodic signal such as a square or sinusoidal wave. The signal is mapped directly to the driving voltage of an electrostatic display, or modulates the high-frequency carrier signal of an ultrasonic display. Bau et al.~\cite{Bau10} have for example shown that varying the frequency and amplitude of an electrovibrating signal can evoke natural sensations such as touching wood, bumpy leather, paper, a painted wall, or rubber, as well as surface properties such as stickiness, smoothness or pleasantness.

While effective and expressive, time-varying periodic signals have limited realism due to their lack of spatial consistency. Unlike physical textures, a time-varying virtual texture's frequency remains fixed no matter how fast the finger moves against it. To increase realism, friction patterns can instead be registered to spatial features of a surface~\cite{Bau12,US9196134}. A realistic, spatially-invariant grating can for example be produced by activating the friction display as a function of finger position instead of time. The improvements in realism are particularly important for large tactile patterns, but less noticeable for fine patterns for which spatial discrepancies are difficult to perceive~\cite{Bau12}. Alternatively, the frequency of the friction signal can be modulated based on the velocity of the finger motion, thereby approximating a spatially-invariant texture~\cite{US9196134}.

Data-driven approaches have also been used to generate realistic textures. Friction profiles are then modeled from physical textures using measurements made by instrumented probes or tribometers. Jiao et al.~\cite{Jiao18} developed a data-driven algorithm for rendering virtual textures on electrostatic friction displays. They acquired force and position data while a finger was sliding on 10 different fabric samples to modulate the voltage applied to a touch screen based on the estimated friction coefficients. They conducted psychophysical experiments and showed that the virtual textures were perceptually similar to the corresponding real textures. Osguei et al.~\cite{osgouei2018inverse} proposed an inverse neural network model to learn from recorded real textures. They showed that the voltage signal for electrostatic actuation can be estimated by this model to render virtual textures on touch screen that resemble the real ones. This is one effective technique for data-driven rendering by compensating for the nonlinear dynamics of an \revc{electrovibration} display \cite{Osgouei2020}. 
Messaoud et al.~\cite{Messaoud16} used multi-level feature extraction to model the friction profiles of two fabrics, twill and velvet. Camillieri et al.~\cite{Camillieri18} evaluated simulations of these fabric on an ultrasonic friction display and found the reproduction of the coefficient of friction to be accurate, but both similarities and differences in the vibrations induced in the finger. Similarly, brain activation was similar for real and virtual twill fabrics, but there were important differences for velvet. 

Wu et al.~\cite{wu2017tactile} proposed a rendering method by electrovibration for image-based textures, based on the estimation of local gradients. The gradients were used to map the frequency (texture granularity) and amplitude (texture height) of the voltage signal applied to the touch screen. Their results showed that modulating frequency further improved tactile realism of virtual textures.

\revn{Despite these advances, tactile rendering algorithms have yet to reach the realism necessary to simulate a wide range of textures. This is likely due to the limitations in current actuation technology and our understanding of tactile perception, which are further discussed in Section \ref{discussion}}

%
% Note to SEUNGMOON: we have decided earlier not to cover any pen/stylus like device
%It should be noted that research on virtual texture rendering on a touchscreen using an active vibration actuator has also been very active. In this approach, providing vibrotactile stimuli of uniform intensities over the entire surface is extremely challenging because of vibration propagation and interference, even if multiple actuators are used. For this reason, the prevalent approach has been using a stylus equipped with a vibration actuator for surface interaction. Under this setup, a large number of texture modeling and rendering methods have been developed. For their review, we refer to \cite{Culbertson2014,Hassan2017,Osgouei2020}.
%

%%%
\subsubsection{Rendering of Virtual Shapes and Surface Features}

In a pilot study, Xu et al.~\cite{Xu11} attempted to display raised dots by electrovibration to form Braille cells, but found the results to be difficult or impossible to read. Attempts to identify three simple geometric shapes (circle, square or triangle) rendered by electrovibration similarly resulted in \revc{a low success rate} of 56\% ~\cite{Xu11}. The difficulty may be largely due to the fact that electrovibration cannot produce a convincing sensation of brushing over a clear edge. Currently, friction modulation produces the same sensation over the entire touch surface. The lack of distributed stimuli under the fingerpad precludes the simulation of a moving edge, and is a likely explanation for the difficulty of identifying shapes with precision.

Despite these technological drawbacks, Kim et al.~\cite{Kim13} proposed an algorithm to render 3D features (or bumps) by electrovibration. The algorithm is based on the observation that we can "create a perception of 3D tactile features on a flat touch surface by manipulating only lateral forces, such as friction." The algorithm maps the gradient of a depth map to the friction display, and normalizes the friction output based on experimental measurements so that the amplitude changes linearly. Their experimental results show that 3D bumps can be perceived by this approach. Osgouei et al.~\cite{Osgouei17} used a similar gradient-based algorithm with electrovibration and asked participants to identify simple geometric patterns such as bumps and holes. They found that participants were unable to do so without contextual information, but had moderate success  when this information was given.

\revn{Ware et al.~\cite{Ware14} have also shown the importance of reducing the need for spatial integration with salient, highly localized features such as tactile edges formed by abrupt changes in friction. Chubb et al.~\cite{Chubb2010} finally demonstrated that surface haptics technologies able to produce net tangential forces (see Section~\ref{section:net-tangential-force}) can facilitate contour following by producing forces orthogonal to finger motion.}

\subsection{User Interface Design}

In this section, we will see how surface haptics can be effectively integrated in user interfaces. We will first survey the tools available to help with the design of user interfaces with surface haptics. We will then go over the specifics of surface haptics by friction modulation that shape how it can be used in interface design. We will then discuss the performance gains expected with the addition of surface haptics in target acquisition tasks, as well as the design of virtual controls and widgets with surface haptics.

\subsubsection{Design Tools} 

While the deployment of haptic technologies is known to be greatly aided by the availability of effective tools for the different steps of the design process (e.g., sketching, prototyping and production)~\cite{moussette2012simple,schneider2017haptic,buxton2010sketching}, few such design tools are currently available for surface haptics.

Solutions for sketching friction-based interfaces have been proposed in~\cite{Potier16, Levesque11}. Levesque et al., in work partially discussed in \cite{Levesque11}, experimented with sketches made by chemically etching glass. The etched areas of the glass reduce the perceived friction, thereby simulating the friction reduction produced by ultrasonic actuation. The resulting patterns, however, were found to be misleading since friction modulation cannot reproduce the distributed tactile stimuli under the fingerpad produced by etched patterns~\cite{maclean17lessons}. Potier et al.~\cite{Potier16} similarly proposed printing patterns on cards using an ink-jet printer and layers of sticky ink, thereby creating friction patterns. This sketching solution was found to produce similar sensations to those of a friction display for certain types of textures.

Although few details have been published, some companies have also released development kits for electrovibration that include development tools such as Application Programming Interfaces (API). \revn{It has also been shown that friction patterns can be specified through bitmapped images (e.g.,~\cite{Mullenbach14}), greatly facilitating experimentation using creative tools for graphic design.}

\setlength{\extrarowheight}{3pt}
\begin{table*}[!ht]
 \caption{Use of current technlogies in applications on haptic surfaces.}
\label{tab:HM}
\centering
\begin{tabularx}{\textwidth}{@{}m{2.5cm}||>{\centering\arraybackslash}m{3cm}CC>{\centering\arraybackslash}m{2.5cm}>{\centering\arraybackslash}m{2cm}CC@{}}
& Normal vibration & Pulses & Lateral vibration & Electrostatic  & Ultrasonic & Driving force & Asymetric friction\\
\toprule
Virtual Texture & \cite{Osgouei2020} & & \cite{wiertlewski:2011}& \cite{Bau10, Bau12,US9196134, Jiao18, osgouei2018inverse, wu2017tactile} & \cite{Messaoud16, Camillieri18} & \\
\hline
Virtual Shapes  & & & & \cite{Kim13, Osgouei17} & & & \cite{Mullenbach2012}\\
\hline
Target Acquisition & & & & \cite{Zhang15} & \cite{Levesque11, Casiez11} & \\
\hline
Buttons & \cite{nashel2003tactile, brewster2007tactile, hoggan2008crossmodal, park2011tactile, pakkanen2010comparison,chen2011design,lylykangas2011designing,kim2014study,wu2015haptic,ma2015haptic,Sadia19} & & \cite{Banter:2010} & &\cite{saleem2017tactile,tashiro2009realization, monnoyer2017optimal,  monnoyer2016ultrasonic} & \cite{gueorguiev:2018} & \\
\hline
Sliding Controls & & & & & \cite{Levesque11,Levesque12, Emgin19, Giraud13a}   & \\
\bottomrule
\end{tabularx}
\end{table*}

\subsubsection{Designing for Friction Modulation} 

Designing user interfaces with friction modulation requires careful consideration due to two current limitations of the technology: (1) friction feedback is only felt while the finger slides against the surface and (2) friction modulation is felt identically over the entire surface.

The first limitation implies that friction modulation typically cannot produce feedback for touch and touch-and-hold interactions~\cite{Bau10}, which require active feedback such as that produced by vibrotactile actuators. This is problematic since these interactions are very common in touch interfaces, e.g., virtual buttons and keyboards that greatly benefit from haptic confirmation feedback. A possible solution consists of combining vibrotactile feedback and friction modulation in the same display, which increases cost and complexity. Alternatively, user interfaces can be redesigned to maximize sliding interactions, e.g. by replacing buttons with sliding toggles~\cite{Levesque11,US9330544}.

The second limitation imposes constraints on the resolution of spatial effects and support for multi-touch interactions. As previously discussed, friction modulation cannot produce the distributed sensation of an edge moving under the fingertip with current technology. Similarly, friction modulation produces the same haptic feedback on all sliding fingers when performing a multi-touch gesture such as a two-finger pinch or rotation. Bau et al.~\cite{Bau10} \revc{proposed} avoiding this limitation by using anchored gestures or two-handed asynchronous manipulations, in which only one finger moves against the surface and feels the feedback. An anchored variant of a rotation gesture, for example, would leave the thumb fixed and rotate the index finger~\cite{Bau10}. This interaction technique was validated experimentally with positive results by Emgin et al.~\cite{Emgin19}. A two-handed asynchronous manipulation similarly consists of an interaction in which one hand is fixed, while the other moves~\cite{Bau10}. It should be noted that these interaction techniques take advantage of the passive nature of friction modulation, and would not be possible with vibrotactile feedback. 

% \cite{Bau10} lists several advantages of electrovibration: no moving part, uniform feedback, scales to any size or shape of surfaces, etc.

%%%
\subsubsection{Target Acquisition}

Pointing at a target is a fundamental task in human-computer interaction that has been extensively studied. The movement time to a target can generally be modelled using Fitt's law, which relates the movement time to the difficulty of the task (size and distance of target) and properties of the interaction modality~\cite{soukoreff2004towards}. The effect of surface haptics on target acquisition has been studied, with results suggesting a benefit of the additional feedback.

Levesque et al.~\cite{Levesque11} investigated the effect of ultrasonic friction modulation on target acquisition for a dragging task. Increasing the friction on a target was shown to improve performance, provided that a single target is displayed. In the presence of spurious targets (distractors), variable friction no longer improved performance but was not detrimental. A supplemental study found evidence for a physical slowdown of the finger with high-friction targets that may explain the performance improvements. Zhang and Harrison~\cite{Zhang15} performed a similar experiment with electrovibration and found a performance improvement of 7.5\% when the entire target is filled with a tactile texture.

Casiez et al.~\cite{Casiez11} similarly studied the effect of ultrasonic friction modulation on an indirect target acquisition task through a haptic trackpad. The results suggest an improvement of close to 9\% in targeting performance in the absence of distractors, and similar performance with distractors. Unlike~\cite{Levesque11}, the effect is believed to be due to information feedback rather a mechanical effect of the friction.

%%%
\subsubsection{Virtual Controls and Widgets}

Virtual controls and widgets are commonly used on touch surfaces but lack the haptic feedback associated with physical controls. This lack of haptic feedback is known to reduce performance and task precision~\cite{Tory13} and often forces users to focus their visual attention on the virtual controller. Tangible controls are sometimes used to restore this haptic feedback with a removable physical object (e.g., \cite{hilliges2007photohelix, weiss2009slap}). SLAP widgets~\cite{weiss2009slap}, for example, have been shown to outperform their virtual counterparts in terms of task completion time and accuracy. Tangible controls, however, reduce flexibility, cause occlusion, and present many other drawbacks due to their physicality. Surface haptics offers many of the advantages of physical or tangible controls, without these practical limitations. The use of surface haptics for two key controls, buttons and sliding widgets, will be discussed next.

\textbf{Buttons.} Buttons are ubiquitous both in physical and digital interfaces. Pressing a physical button produces a distinct response profile, with a satisfying and informative click on activation. In contrast, virtual buttons on touchscreen interfaces typically offer limited haptic feedback, if any. 

Augmenting a virtual button with haptic feedback has typically been done with low-cost vibration motors and voice coils~\cite{nashel2003tactile, brewster2007tactile, hoggan2008crossmodal, park2011tactile} or piezoelectric actuators~\cite{pakkanen2010comparison,chen2011design,lylykangas2011designing,kim2014study,wu2015haptic,ma2015haptic,Sadia19}. Park et al.~\cite{park2011tactile} investigated the effect of parameters such as the amplitude, duration, carrier signal, envelope function, and type of actuator on the rendering of a virtual button on a mobile device. Their results suggest that a short rise time and short duration are preferred for a pleasant button click. Similarly, Chen et al.~\cite{chen2011design} found that frequencies of 500~Hz and 125~Hz result in key clicks that are perceived as crisp and dull, respectively. Lylykangas et al.~\cite{lylykangas2011designing} found significant effects of time delay and vibration duration when augmenting clicks with piezoelectric vibrations.

More recently, Sadia et al.~\cite{Sadia19} displayed vibrotactile feedback through a touch surface with piezoelectric actuators to simulate the feeling of physical buttons. They first recorded and analyzed the force, acceleration, and voltage data from the interactions of twelve participants with three different physical buttons: latch, toggle, and push buttons. Then, a button-specific vibrotactile stimulus was generated for each button by modulating the recorded acceleration signal with the resonance frequency of the surface. In their user experiments, participants were able to match the three digital buttons with their physical counterparts with a success rate of 83\%.

Although button clicks are most commonly produced with vibration feedback, it has been shown that the effect can be simulated by friction modulation on ultrasonic devices~\cite{saleem2017tactile,tashiro2009realization, monnoyer2017optimal,  monnoyer2016ultrasonic}. Tashiro et al.~\cite{tashiro2009realization} reproduced the sensation of buckling and restitution of a physical button using a Langevin-type ultrasonic transducer and explained the sensation by the momentary finger slippage as friction is reduced. Monnoyer et al.~\cite{monnoyer2016ultrasonic}, on the other hand, explains the clicking sensation by the accumulation and sudden release of stress in the fingerpad. The sensation of click has also been optimized based on \revc{the finger impedance}~\cite{monnoyer2017optimal} and the rate of change in normal force~\cite{Saleem18}.

\textbf{Sliding Controls.} While simulating button clicks may be possible with certain variants of friction modulation, this type of surface haptics is typically best used to augment controls such as knobs and sliders that require relative motion between the finger and the interactive surface. 

Emgin et al.~\cite{Emgin19} rendered a virtual knob with tactile feedback on the HapTable, a large electrostatic display. Friction feedback such as detents was produced as users turned the knob to select items in a menu. While performance improvements could not be demonstrated, the haptic feedback was found to significantly improve the subjective experience of the participants. Giraud et al.~\cite{Giraud13a} similarly implemented a haptic knob with ultrasonic friction modulation and found improvements in task accuracy when selecting an item in a scrolling list.

As further discussed below, Levesque et al.~\cite{Levesque11,Levesque12} also experimented with the design of sliding controls with ultrasonic friction modulation on touchscreens. They first designed and evaluated a number of sample applications using sliding gestures (such as moving the wheels of an alarm clock)~\cite{Levesque11}, before evaluating the feasibility of several scrolling interactions~\cite{Levesque12}.

%%%%%%%%%%%%%%%%%%%%%%%%%%%%%%%%%%%%%%%%%%%%%%%%%%%
\subsection{Applications}

Several researchers have explored applications of surface haptics, and particularly the novel possibilities offered by friction modulation. 

Bau et al.~\cite{Bau10} proposed a wide range of applications of surface haptics by vibration or friction modulation, with the expectation of better results with electrovibration due to its frequency range and uniform response over a surface. The applications include the simulation of realistic tactile effects (e.g., the friction between a brush and a canvas), non-visual information layers (e.g., a tactile overlay indicating radiation intensity over the image of a star), augmented GUI widgets (e.g., sliders or scrolling lists with tactile feedback), direct manipulation (e.g., friction on drag \& drop operations), or rubbing interactions (e.g., erasing by repeated brushing). Some of these application concepts were implemented on an electrovibrating surface, such as feeling different textures, dragging a race car along a track, and editing a picture by rubbing.

Levesque et al.~\cite{Levesque11} explored the design space of ultrasonic friction reduction for touchscreens. They designed four simple applications demonstrating several use cases for friction: an alarm clock, a file manager, a game and a text editor. The effects included clicks when turning the wheels of the alarm clock, resistance when dragging a file over a folder, impacts as a ball is hit in the game, and a "pop-through" sensation when displacing words in the text editor. The resulting user experiences were evaluated quantitatively and qualitatively by having participants briefly interact with the sample applications with and without haptic feedback. The results suggest that the tactile feedback had a positive impact on enjoyment, engagement and the sense of realism.

In follow-up work, Levesque et al.~\cite{Levesque12} further explored the applications of friction modulation for scrolling interactions on touchscreens. They developed five scenarios exemplifying interaction styles and applications such as scrolling in a webpage or adjusting a numerical value on a slider. Five experiments were run to understand the feasibility of the different required features, such as being able to identify and count detents, and perceive their density. The results were recommendations for design based on the strengths and limitations that were identified.

Although few details are publicly available, some companies have also developed demonstrators for surface haptics. Levesque et al.~\cite{US9330544}, for example, \revc{describe} a demonstration application inspired by illustrated children's books with tactile features, such as fabrics, fur and moving parts. The virtual book contains five interactive pages exemplifying the strengths of electrovibration: four animals with different textures, from soft fur to scales; a block of ice that can be rubbed away, revealing a monster; a suspended car that can be released by cutting or sawing ropes; a sliding character that brushes against the page as it is moved; and a safe that can be opened by turning a dial. Of particular note, activation of the feedback is controlled by pulling on a ribbon instead of pressing on a button, thereby allowing friction feedback to be produced. 

Mullenbach et al.~\cite{Mullenbach14} explored applications of ultrasonic friction modulation for affective communication. Three applications were implemented and evaluated: a text messaging application in which messages can be augmented with one of 24 friction patterns; an image sharing application in which an image can be annotated with friction patterns; and a virtual touch application in which remote partners can feel each other's traces against a touchscreen. The results suggest that the participants, which included strangers and intimate couples, viewed surface haptics as an effective way to communicate emotional information to a remote partner.

Applications of surface haptics for the visually impaired have also been explored. Xu et al.~\cite{Xu11} experimented with electrovibration to produce Braille and tactile graphics. Since electrovibration cannot produce a pattern smaller than the fingertip, Braille was presented through frequency modulation, temporal mapping, or enlargement. The results were found to be difficult to read. Attempts to display simple geometrical shapes (circle, square and triangle) similarly resulted in relatively low performance (56\% identification rate), likely due to the difficulty of representing sharp edges with electrovibration. Israr et al.~\cite{Israr12} similarly used electrovibration for sensory substitution, allowing visually impaired users to aim a touchscreen at a scene and feel its content through friction patterns. The device was tested by having two visually impaired participants search for an object in a room, with positive results. Lim et al. \cite{Lim2019} designed and prototyped a mobile application, called TouchPhoto, that allows a visually-impaired user to take photographs independently while recording auditory tags that are useful for recall of the photographs' content. TouchPhoto also provides a system with which the user can perceive the main landmarks of a person's face in the photograph by touch using an electrovibration display. However, a user study showed that the effectiveness of haptics had considerable room for improvement.

Bau et al.~\cite{Bau12} expanded the use of electrovibration beyond touchscreens and explored applications of tactile feedback for real-world objects. The work was positioned as a form of tactile augmented reality, in which the tactile feeling of real-world objects can be augmented with virtual tactile effects. Sample applications include augmented surfaces (e.g., tactile feedback on a projected wall display), see-through augmented reality with tactile feedback on real-world objects, and tactile feedback for interactions with tangibles. Further concepts were also proposed, such as adding tactile textures to physical prints, communicating personalized, private messages in public touch displays, and assistive applications (e.g., guiding a visually impaired person with a tactile guide against a wall).

%As can be seen for the description above, many applications of surface haptics have been conceptually explored, some have been implemented, and few have been formally evaluated through user experiments. Much work remains to further explore, optimize and validate these novel user experiences, and to understand how to eliminate or get around the limitations of current surface haptic technologies.
\section{DISCUSSION and CHALLENGES AHEAD} \label{discussion}
%5 Ahead\\5.2. Future Outlook\\ 
%Lead: VL, Contributors: SC, FG, CB
\subsection{Challenges} 
Currently, there are three prominent actuation methods for displaying tactile feedback through a touch surface; vibrotactile, electrostatic, and ultrasonic. Each actuation method has its own advantages and disadvantages. However, independent of the actuation method being used, there are a number of features \textit{desired} for surface haptic displays as listed below. Implementation of some is highly challenging with the technology available today, but necessary to make surface haptics displays reach the mass market in the near future.

\textbf{High Transparency.} The tactile surface should be transparent to allow visual information to be displayed to the user in tandem with tactile feedback. This puts some constraints on material properties of the touch surface as well as the type of actuators and actuation methods used for tactile feedback. For example, it is known that a brittle and transparent glass surface vibrates less than a ductile and non-transparent metal surface of the same size and shape. Hence, more powerful actuators are required to display vibrotactile tactile feedback on \revc{a} glass surface with sufficient amplitude. Partially for the same reason, the zones where the friction reduction is achieved are limited in ultrasonic actuation.

\textbf{Large Tactile Interaction Area.} The interaction area should be sufficiently large to display effective tactile feedback to the \revc{users as they interact} with digital data. However, achieving a large tactile area is not always possible due to the limitations of the current technology. For example, ultrasonic actuation of a touch surface at a certain resonance frequency leads to a limited number of effective zones for friction modulation due to the certain mode shape at that frequency based on the dimensions and material properties of the surface. Similarly, the number of potential problems increases with the size of capacitive touchscreens used for electrostatic actuation due to the difficulties in manufacturing them. Large capacitive touch screens may cause non-uniform distribution of charges on its surface as a result of the variations in the thickness of conductive and insulator layers and also the non-visible cracks in the insulator layer.   

\textbf{Simultaneous Tactile Feedback Displayed in Different Directions.} It is desired to display tactile feedback in the directions normal and tangential to the tactile surface simultaneously. However, it is not always possible to achieve tactile effects in all directions using a single actuation technique today. For example, it is currently not possible to generate tactile effects in the normal direction using electrostatic actuation while vibrotactile actuation is typically preferred over the others for this purpose.

\textbf{Simultaneous Tactile Feedback Stimulating Different Receptors.} The current actuation technology is limited to the stimulation of rapidly adapting mechanoreceptors mostly. For example, due to the nature of electrovibration, it is not possible to feel bumps or edges when the finger is stationary. In addition to the four major mechanoreceptors mentioned in Section 3, \revc{the} human finger is equipped with thermoreceptors, nociceptors, and chemoreceptors. Among those, there are a limited number of studies on stimulating thermoreceptors by modulating the temperature of a surface while stimulating mechanoreceptors. \revn{For example, Guo et al. \cite{guo2019effect} investigated the effect of temperature on tactile perception of stimulus displayed by electrovibration and reported that the absolute detection threshold decreases with increasing temperature.}

\textbf{Low Power Consumption.} The tactile surface should not require a significant energy to actuate. One of the challenges with the current surface haptics displays is the high energy requirements, which hinders them to reach the mass market. For example, both the ultrasonic and electrostatic actuation techniques require significantly high voltages to operate, compared to the voltage requirements of low-cost vibration motors typically used in mobile devices today. The frequency of stimulation is another factor affecting the energy cost. For example, in order to display virtual textures, a low frequency texture signal is typically modulated with a high frequency carrier signal close to the sensitive frequencies of \revc{the} human vibrotactile system to boost the tactile effect. While amplitude modulation is beneficial for effective rendering of complex tactile signals, it is disadvantageous in terms of the energy cost.

\textbf{Easy Integration and Compact Design.} It should be easy to integrate a tactile surface into current devices such as \revc{mobile} phones and tablets. For those systems, for example, the number of moving parts needs to be reduced to \revc{prevent} dirt and dust \revc{from penetrating} into the devices. The geometry and material properties of \revc{the} tactile surface, the location of actuators attached to it, and the housing in which \revc{it} is placed \revc{all have} significant influence on the ease of integration, but also the tactile feedback capacity of the device. For example, in the case of \revc{vibrotactile} and ultrasonic actuation, the boundary conditions influence the out of plane and in-plane resonance frequencies and mode shapes of the tactile surface. In the case of electrostatic actuation on the other hand, where there is no mechanical vibration, the thickness of insulator layer has \revc{been} shown to affect the magnitude of friction force displayed to the user~\cite{sirin2019electroadhesion}. While it is possible to reduce the applied voltage by using a thinner insulator, the thickness is still limited by the breakdown voltage. The breakdown voltage is the maximum voltage difference that can be applied across the conductive layers (ITO and the finger) before the insulator collapses and conducts current.

\textbf{Realistic Tactile Effects.} Not only the richness of tactile effects, but the flexibility in generating them and their realism are also highly important. Again, due to the limitations in current technology, it is difficult to achieve all together. For example, in texture rendering, it is not possible to display texture topography directly by modulating friction in the tangential direction using ultrasonic or electrostatic actuation. As a result, tactile perception of virtual textures displayed on touch surfaces via those methods is expected to be different than tactile perception of real ones \cite{Isleyen19}, \cite{saleemtactile}. Moreover, the range for the modulated friction coefficient is limited in both methods though their coupling results in a larger range \cite{vezzoli2015physical}. Even if we have a larger bandwidth, the realistic rendering of virtual textures on touch surfaces will be still challenging since the textures in nature are complex and come in various forms. \revn{Recently, a tri-modal tactile display was developed by integrating electrostatic, ultrasonic, and vibrotactile actuation methods \cite{liu2020tri}. The authors claim that this approach leads to significant improvements both in recognition rate and tactile perception of tactile images.}  

\textbf{Localized, \revn{Distributed}, and Multi-Finger Interactions.} Future surface haptics applications will include multiple tactile stimuli displayed simultaneously or consecutively to \revc{a} single finger or multiple fingers through a touch surface. However, there are technological challenges in displaying localized tactile stimuli through a touch surface in multi-finger applications. For example, in the case of vibrotactile feedback in the normal and tangential directions, it is highly difficult to control the propagation of the vibration waves to create localized effects\revc{,} though some solutions have been already suggested in the literature as discussed in Section~\ref{localization}. These solutions\revc{,} however\revc{,} require multiple actuators with independent control. The challenge is to be able to adapt them to the localization of the tactile stimuli, to the surface geometry, and to the number of fingers touching the surface. It is therefore essential to design control electronics that can address the aforementioned issues at low cost, yet ensuring fast response.

On the other hand, in the case of electrovibration, current design and manufacturing steps of multi-finger capacitive touch screens (i.e.\revc{,} projected capacitive) introduce some difficulties. Today, most of the research groups working on surface haptics \revc{use} a single-touch capacitive screen (i.e.\revc{,} surface capacitive) to avoid those difficulties. A surface capacitive touchscreen \revc{uses} a single layer of conductive ITO with a simple electrode pattern on top of it and passing through its edges. Hence, the voltage applied to the conductive ITO layer generates uniform tactile effects \revc{over its surface}. If the finger position is tracked by means of the touch screen itself or an external sensor such as an IR frame, then different tactile effects can be displayed to the user based on the acquired finger position, creating an illusion of localized tactile feedback. However, this illusion breaks down quickly as soon as the user contacts the surface with more than one finger. In \revc{a} true multi-finger case, each finger has to be stimulated with a different and independent tactile signal. In order to demonstrate the concept of true multi-touch haptic interactions\revc{,} for example, Emgin et al. ~\cite{Emgin2015_HardwareDemo} developed a custom solution. They ablated a surface capacitive touch screen using a UV laser to pattern a grid of conductive indium tin oxide (ITO) cells (called haptiPads by the authors) and thin ITO wires carrying electric current to the cells for independent electrostatic actuation. This approach leads to localized and multi-finger haptics, but the process of constructing haptiPads is tedious and error-prone in a university laboratory. Perhaps, a better alternative is to display tactile effects through a projected capacitive screen. In fact, surface capacitive touch screens are already replaced by projected-capacitive touchscreens in the market since the latter allows simultaneous sensing of multi-finger contacts. Projected-capacitive screens \revc{use} one or more etched ITO layers forming multiple horizontal (X) and vertical (Y) electrodes in various patterns to detect multi-finger contact interactions. Obviously, when a voltage applied to a single horizontal/vertical electrode, electrovibration is generated along a thin row/column on the touch screen. In order \revc{to} create localized stimulation, Haga et al.~\cite{haga2015electrostatic} used \revc{the} beat phenomenon. When two sinusoidal signals with frequencies of $f_1$ and $f_2$ are superposed, the resulting “beat” signal will have major frequency components at $(f_1+f_2)/2$ and $(f_1-f_2)/2$. They applied AC voltages with frequencies of 1000 Hz and 1240 Hz to an X and Y electrodes to create an oscillating electrostatic force at frequencies of 2240 Hz (2 x 1120 Hz) and 240 Hz (2 x 120 Hz). The first frequency is outside the detectable range of human vibrotactile sensing while the latter is optimal.

\revn{If the number of X and Y electrodes are sufficiently high and can be actuated individually, then we can talk about \textit{distributed} surface haptics. Unfortunately, current surface haptics displays lack a distributed presentation. This means that an edge smaller than the size of a finger cannot be displayed when it is touched with one finger.}

\subsection{Future Outlook} 
\revc{Nowadays}, touch surfaces are an integral part of our mobile phones, smart watches, tablets, laptops, ATMs, tabletops, vending
machines, electronic kiosks, and car navigation systems. 
These surfaces allow us to directly manipulate digital content via
touch gestures. However, the lack of sophisticated tactile feedback displayed through touch surfaces results in a decrease in user experience quality and even task performance. Moreover, it is also important to realize that most of our sensory communication with the \revc{ aforementioned} electronic devices today is through visual and auditory channels, which are highly overloaded. Alternatively, some of the information can be displayed through the haptic channel in order to alleviate the perceptual and cognitive load of the user. In this regard, touch surfaces appear to be the right user interface and the research on surface haptics \revc{could have a significant return on investment} in the near future. The largest impact is naturally expected to be in mobile devices. In this domain, gesture-based interaction with touch surfaces is already the primary user interface and \revc{makes} it easier to access and manipulate digital data. With the integration of tactile feedback \revc{in} existing gestures such as tap, slide, click, drag, pinch, twirl, and swipe, our performance and confidence in activating, selecting, moving, zooming, rotating, and adjusting digital content is expected to improve. In addition, tactile feedback on mobile devices will enable the visually impaired and blind to access digital content. In fact, even sighted people may rely on tactile feedback in difficult ambient lighting conditions.   

However, we anticipate that surface haptics in the future will not be restricted to the touch surfaces of electronic devices only but will be available in any flat, curved, and flexible physical surface made of hard or soft material having some form of embedded computational capability. \revn{In particular, emerging new material technologies can bridge the electronic and mechanical domains (see a recent review in \cite{biswas2019emerging}), enabling more sophisticated surface displays}.

%\input{introduction.tex}
%\input{review.tex}
%\input{discussion.tex}

%
% ACKNOWLEDGEMENTS
%
\section*{Acknowledgment}
C.B. acknowledges the financial support provided by The Scientific and Technological Research Council of Turkey (TUBITAK) under the contract no. 117E954. F.G. acknowledges the support of IRCICA, USR CNRS 3380. \revc{V.L. acknowledges the support of the Discovery Grant program of the Natural Sciences and Engineering Research Council (NSERC) of Canada.}

% Can use something like this to put references on a page
% by themselves when using endfloat and the captionsoff option.
\ifCLASSOPTIONcaptionsoff
  \newpage
\fi

% trigger a \newpage just before the given reference
% number - used to balance the columns on the last page
% adjust value as needed - may need to be readjusted if
% the document is modified later
%\IEEEtriggeratref{8}
% The "triggered" command can be changed if desired:
%\IEEEtriggercmd{\enlargethispage{-5in}}

%
% REFERENCES
%
\bibliographystyle{IEEEtran}
\bibliography{IEEEabrv,references,./SChoi/schoibib_old}

% Generated by IEEEtran.bst, version: 1.14 (2015/08/26)
\begin{thebibliography}{100}
\providecommand{\url}[1]{#1}
\csname url@samestyle\endcsname
\providecommand{\newblock}{\relax}
\providecommand{\bibinfo}[2]{#2}
\providecommand{\BIBentrySTDinterwordspacing}{\spaceskip=0pt\relax}
\providecommand{\BIBentryALTinterwordstretchfactor}{4}
\providecommand{\BIBentryALTinterwordspacing}{\spaceskip=\fontdimen2\font plus
\BIBentryALTinterwordstretchfactor\fontdimen3\font minus
  \fontdimen4\font\relax}
\providecommand{\BIBforeignlanguage}[2]{{%
\expandafter\ifx\csname l@#1\endcsname\relax
\typeout{** WARNING: IEEEtran.bst: No hyphenation pattern has been}%
\typeout{** loaded for the language `#1'. Using the pattern for}%
\typeout{** the default language instead.}%
\else
\language=\csname l@#1\endcsname
\fi
#2}}
\providecommand{\BIBdecl}{\relax}
\BIBdecl

\bibitem{Jansen10}
Y.~Jansen, ``Mudpad: Fluid haptics for multitouch surfaces,'' in \emph{CHI '10
  Extended Abstracts on Human Factors in Computing Systems}, ser. CHI EA '10,
  2010, pp. 4351--4356.

\bibitem{Craig12}
``User interface system and method,'' Patent PCT/US2011/057\,175, apr 26, 2012.

\bibitem{Follmer13}
S.~Follmer, D.~Leithinger, A.~Olwa, A.~Hogge, and H.~Ishii, ``inform: Dynamic
  physical affordances and constraints through shape and object actuation,'' in
  \emph{Proceedings of the 26th Annual ACM Symposium on User Interface Software
  and Technology}, ser. UIST '13, 2013, pp. 417--426.

\bibitem{Hoshi10}
T.~{Hoshi}, M.~{Takahashi}, T.~{Iwamoto}, and H.~{Shinoda}, ``Noncontact
  tactile display based on radiation pressure of airborne ultrasound,''
  \emph{IEEE Transactions on Haptics}, vol.~3, no.~3, pp. 155--165, July 2010.

\bibitem{Kyung09}
K.~{Kyung} and J.~{Lee}, ``Ubi-pen: A haptic interface with texture and
  vibrotactile display,'' \emph{IEEE Computer Graphics and Applications},
  vol.~29, no.~1, pp. 56--64, Jan 2009.

\bibitem{Yao:2010}
H.-Y. Yao and V.~Hayward, ``Design and analysis of a recoil-type vibrotactile
  transducer,'' \emph{The Journal of the Acoustical Society of America}, vol.
  128, no.~2, pp. 619--627, 2010.

\bibitem{Zhao2015_WHC}
S.~Zhao, A.~Israr, and R.~L. Klatzky, ``Intermanual apparent tactile motion on
  handheld tablets,'' in \emph{Proceedings of the IEEE World Haptics
  Conference}, 2015, pp. 241--247.

\bibitem{Park2018}
G.~Park and S.~Choi, ``Tactile information transmission by 2d stationary
  phantom sensations,'' in \emph{Proceedings of the 2018 CHI Conference on
  Human Factors in Computing Systems}, New York, New York, USA, 2018, pp.
  1--12, paper 258.

\bibitem{Hudin:2018}
C.~Hudin and S.~Pan{\"e}els, ``Localisation of vibrotactile stimuli with
  spatio-temporal inverse filtering,'' in \emph{Haptics: Science, Technology,
  and Applications}, D.~Prattichizzo, H.~Shinoda, H.~Z. Tan, E.~Ruffaldi, and
  A.~Frisoli, Eds.\hskip 1em plus 0.5em minus 0.4em\relax Cham: Springer
  International Publishing, 2018, pp. 338--350.

\bibitem{Woo:2015}
J.-H. Woo and J.-G. Ih, ``Vibration rendering on a thin plate with actuator
  array at the periphery,'' \emph{Journal of Sound and Vibration}, vol. 349,
  pp. 150 -- 162, 2015.

\bibitem{Enferad:2018}
E.~{Enferad}, C.~{Giraud-Audine}, F.~{Giraud}, M.~{Amberg}, and
  B.~{Lemaire-Semail}, ``Differentiated haptic stimulation by modal synthesis
  of vibration field,'' in \emph{2018 IEEE Haptics Symposium (HAPTICS)}, March
  2018, pp. 216--221.

\bibitem{Bai:2011}
M.~R. Bai and Y.~K. Tsai, ``Impact localization combined with haptic feedback
  for touch panel applications based on the time-reversal approach,'' \emph{The
  Journal of the Acoustical Society of America}, vol. 129, no.~3, pp.
  1297--1305, 2011.

\bibitem{Hudin:2013}
C.~{Hudin}, J.~{Lozada}, and V.~{Hayward}, ``Localized tactile stimulation by
  time-reversal of flexural waves: Case study with a thin sheet of glass,'' in
  \emph{2013 World Haptics Conference (WHC)}, April 2013, pp. 67--72.

\bibitem{wiertlewski:2011}
M.~{Wiertlewski}, J.~{Lozada}, and V.~{Hayward}, ``The spatial spectrum of
  tangential skin displacement can encode tactual texture,'' \emph{IEEE
  Transactions on Robotics}, vol.~27, no.~3, pp. 461--472, June 2011.

\bibitem{Nara:2001}
T.~{Nara}, M.~{Takasaki}, T.~{Maeda}, T.~{Higuchi}, S.~{Ando}, and S.~{Tachi},
  ``Surface acoustic wave tactile display,'' \emph{IEEE Computer Graphics and
  Applications}, vol.~21, no.~6, pp. 56--63, Nov 2001.

\bibitem{Biet:2007}
M.~{Biet}, F.~{Giraud}, and B.~{Lemaire-Semail}, ``Squeeze film effect for the
  design of an ultrasonic tactile plate,'' \emph{IEEE Transactions on
  Ultrasonics, Ferroelectrics, and Frequency Control}, vol.~54, no.~12, pp.
  2678--2688, December 2007.

\bibitem{Winfield:2007}
L.~{Winfield}, J.~{Glassmire}, J.~E. {Colgate}, and M.~{Peshkin}, ``T-pad:
  Tactile pattern display through variable friction reduction,'' in
  \emph{Second Joint EuroHaptics Conference and Symposium on Haptic Interfaces
  for Virtual Environment and Teleoperator Systems (WHC'07)}, March 2007, pp.
  421--426.

\bibitem{ghenna:2017}
S.~{Ghenna}, C.~{Giraud-Audine}, F.~{Giraud}, M.~{Amberg}, and
  B.~{Lemaire-Semail}, ``Control and evaluation of a 2-d multimodal
  controlled-friction display,'' in \emph{2017 IEEE World Haptics Conference
  (WHC)}, June 2017, pp. 593--598.

\bibitem{Grimnes83}
S.~Grimnes, ``Electrovibration, cutaneous sensation of microampere current,''
  \emph{Acta Physiologica Scandinavica}, vol. 118, no.~1, pp. 19--25, 1983.

\bibitem{Strong70}
R.~M. {Strong} and D.~E. {Troxel}, ``An electrotactile display,'' \emph{IEEE
  Transactions on Man-Machine Systems}, vol.~11, no.~1, pp. 72--79, March 1970.

\bibitem{beebe1995polyimide}
D.~J. Beebe, C.~Hymel, K.~Kaczmarek, and M.~Tyler, ``A polyimide-on-silicon
  electrostatic fingertip tactile display,'' in \emph{Proceedings of 17th
  International Conference of the Engineering in Medicine and Biology Society},
  vol.~2, 1995, pp. 1545--1546.

\bibitem{senseg:2008}
J.~L. Ville~M\"{a}kinen, Petro~Suvanto, ``Interface apparatus for touch input
  and tactile output communication,'' Finland Patent PCT/FI2009/050\,416, nov
  26, 2009.

\bibitem{Chubb2010}
E.~Chubb, J.~Colgate, and M.~Peshkin., ``Shiverpad: a glass haptic surface that
  produces shear force on a bare finger,'' \emph{IEEE Transactions on Haptics},
  vol.~3, no.~3, pp. 189--198, 2010.

\bibitem{ghenna:2017b}
S.~{Ghenna}, E.~{Vezzoli}, C.~{Giraud-Audine}, F.~{Giraud}, M.~{Amberg}, and
  B.~{Lemaire-Semail}, ``Enhancing variable friction tactile display using an
  ultrasonic travelling wave,'' \emph{IEEE Transactions on Haptics}, vol.~10,
  no.~2, pp. 296--301, April 2017.

\bibitem{Ryu2010}
J.~Ryu, ``Perception-based vibration rendering in mobile device,'' Ph.D.
  dissertation, Pohang University of Science and Engineering (POSTECH), 2010.

\bibitem{Choi2013}
S.~Choi and K.~J. Kuchenbecker, ``Vibrotactile display: Perception, technology,
  and applications,'' \emph{Proceedings of the IEEE}, vol. 101, no.~9, pp.
  2093--2104, 2013.

\bibitem{Dhiab:2019}
A.~B. {Dhiab} and C.~{Hudin}, ``Confinement of vibrotactile stimuli in narrow
  plates,'' in \emph{2019 IEEE World Haptics Conference (WHC)}, July 2019, pp.
  431--436.

\bibitem{Aono:2014}
T.~Aono, ``Input apparatus and control method of input apparatus,'' U.S. Patent
  8,830,187, sept 09, 2014.

\bibitem{Kess:2015}
P.~Kessler, D.~C. Patel, J.~A. Harley, B.~W. Degner, N.~A. Rundle, P.~K.
  Augenbergs, N.~Lubinski, K.~L. Staton, and O.~S. Leung, ``Haptic
  electromagnetic actuator,'' U.S. Patent Application 14/404,156, jun 10, 2015.

\bibitem{Mullenbach2012}
J.~Mullenbach, D.~Johnson, J.~Colgate, and M.~Peshkin, ``Activepad surface
  haptic device,'' in \emph{Proceedings of the IEEE Haptics Symposium
  (HAPTICS)}, 2012, pp. 407--414.

\bibitem{imaizuni:2014}
A.~Imaizumi, S.~Okamoto, and Y.~Yamada, ``Friction sensation produced by
  laterally asymmetric vibrotactile stimulus,'' in \emph{Haptics: Neuroscience,
  Devices, Modeling, and Applications}, M.~Auvray and C.~Duriez, Eds.\hskip 1em
  plus 0.5em minus 0.4em\relax Berlin, Heidelberg: Springer Berlin Heidelberg,
  2014, pp. 11--18.

\bibitem{Poupyrev:2002}
I.~Poupyrev, J.~Rekimoto, and S.~Maruyama, ``Touchengine: A tactile display for
  handheld devices,'' in \emph{CHI '02 Extended Abstracts on Human Factors in
  Computing Systems}, ser. CHI EA '02, 2002, pp. 644--645.

\bibitem{Emgin19}
S.~E. Emgin, A.~Aghakhani, T.~M. Sezgin, and C.~Basdogan, ``Haptable: an
  interactive tabletop providing online haptic feedback for touch gestures,''
  \emph{IEEE Transactions on Visualization and Computer Graphics}, vol.~25,
  no.~9, pp. 2749--2762, 2019.

\bibitem{Dai:2012b}
X.~Dai, J.~Gu, X.~Cao, J.~E. Colgate, and H.~Tan, ``Slickfeel: Sliding and
  clicking haptic feedback on a touchscreen,'' in \emph{Adjunct Proceedings of
  the 25th Annual ACM Symposium on User Interface Software and Technology},
  ser. UIST Adjunct Proceedings '12, 2012, pp. 21--22.

\bibitem{Hudin:2015}
C.~{Hudin}, J.~{Lozada}, and V.~{Hayward}, ``Localized tactile feedback on a
  transparent surface through time-reversal wave focusing,'' \emph{IEEE
  Transactions on Haptics}, vol.~8, no.~2, pp. 188--198, April 2015.

\bibitem{Wiertlewski:2013}
M.~Wiertlewski, \emph{Causality Inversion in the Reproduction of
  Roughness}.\hskip 1em plus 0.5em minus 0.4em\relax London: Springer London,
  2013, pp. 45--53.

\bibitem{Hap2u}
``Website of hap2u,'' https://www.hap2u.net/, accessed on November 1st 2019.

\bibitem{takasaki:2019}
H.~Kimura, M.~Takasaki, D.~Yamaguchi, Y.~Ishino, and T.~Mizuno, ``Higher
  resonant frequency of bulk wave ultrasonic transducer for tactile display,''
  in \emph{International Workshop on Piezoelectric Materials and Applications
  in Actuators 2019}, October 2019, pp. 169--170.

\bibitem{dai:2012}
{Xiaowei Dai}, J.~E. {Colgate}, and M.~A. {Peshkin}, ``Lateralpad: A
  surface-haptic device that produces lateral forces on a bare finger,'' in
  \emph{2012 IEEE Haptics Symposium (HAPTICS)}, March 2012, pp. 7--14.

\bibitem{xu:2019}
H.~{Xu}, M.~A. {Peshkin}, and E.~{Colgate}, ``Ultrashiver: Lateral force
  feedback on a bare fingertip via ultrasonic oscillation and
  electroadhesion,'' \emph{IEEE Transactions on Haptics}, pp. 1--1, 2019.

\bibitem{Duong:2019}
Q.~Van~Duong, V.~P. Nguyen, F.~Domingues Dos~Santos, and S.~T. Choi,
  ``Localized fretting-vibrotactile sensations for large-area displays,''
  \emph{ACS Applied Materials \& Interfaces}, vol.~11, no.~36, pp.
  33\,292--33\,301, 2019.

\bibitem{Frisson:2017}
C.~Frisson, J.~Decaudin, T.~P.~A. Ng, P.~Poncet, F.~Casset, A.~Latour, and
  S.~A. Brewster, ``Designing vibrotactile widgets with printed actuators and
  sensors,'' in \emph{Adjunct Publication of the 30th Annual ACM Symposium on
  User Interface Software and Technology}, ser. UIST '17, 2017, pp. 11--13.

\bibitem{yamamoto2006electrostatic}
A.~Yamamoto, S.~Nagasawa, H.~Yamamoto, and T.~Higuchi, ``Electrostatic tactile
  display with thin film slider and its application to tactile telepresentation
  systems,'' \emph{IEEE Transactions on Visualization and Computer Graphics},
  vol.~12, no.~2, pp. 168--177, 2006.

\bibitem{tanvas}
``Website of tanvas,'' https://tanvas.co/, accessed on November 1st 2019.

\bibitem{Goldstein02}
E.~B. Goldstein, \emph{Sensation and Perception}, 6th~ed.\hskip 1em plus 0.5em
  minus 0.4em\relax Pacific Grove, CA, USA: Wadsworth-Thomson Learning, 2002.

\bibitem{Allies1970}
D.~S. Alles, ``Information transmission by phantom sensations,'' \emph{IEEE
  Transactions on Man-Machine Systems}, vol. MMS-11, no.~1, pp. 85--91, 1970.

\bibitem{Jones2006}
L.~A. Jones and S.~J. Lederman, \emph{Human Hand Function}.\hskip 1em plus
  0.5em minus 0.4em\relax New York, NY, USA: Oxford University Press, Inc.,
  2006.

\bibitem{RyuYangKang2009}
D.~Ryu, G.-H. Yang, and S.~Kang, ``{T-Hive}: Vibrotactile interface presenting
  spatial information on handle surface,'' in \emph{Proceedings of the IEEE
  International Conference on Robotics and Automation (ICRA)}, 2009, pp.
  683--688.

\bibitem{YangRyuKang2009}
G.-H. Yang, D.~Ryu, and S.~Kang, ``Vibrotactile display for hand-held input
  device providing spatial and directional information,'' in \emph{Proceedings
  of the World Haptics Conference (WHC)}.\hskip 1em plus 0.5em minus
  0.4em\relax IEEE, 2009, pp. 79--84.

\bibitem{Park2017}
\BIBentryALTinterwordspacing
G.~Park, H.~Cha, and S.~Choi, ``{Attachable and detachable vibrotactile
  feedback modules and their information capacity for spatiotemporal
  patterns},'' in \emph{2017 IEEE World Haptics Conference (WHC)}, no.
  June.\hskip 1em plus 0.5em minus 0.4em\relax IEEE, jun 2017, pp. 78--83.
  [Online]. Available: \url{http://ieeexplore.ieee.org/document/7989880/}
\BIBentrySTDinterwordspacing

\bibitem{Park2019a}
\BIBentryALTinterwordspacing
------, ``{Haptic Enchanters: Attachable and Detachable Vibrotactile Modules
  and Their Advantages},'' \emph{IEEE Transactions on Haptics}, vol.~12, no.~1,
  pp. 43--55, jan 2019. [Online]. Available:
  \url{https://ieeexplore.ieee.org/document/8419276/}
\BIBentrySTDinterwordspacing

\bibitem{KimSY2009}
S.-Y. Kim, J.-O. Kim, and K.~Y. Kim, ``Traveling vibrotactile wave - a new
  vibrotactile rendering method for mobile devices,'' \emph{IEEE Transactions
  on Consumer Electronics}, vol.~55, no.~3, pp. 1032--1038, 2009.

\bibitem{Kim2012}
S.-Y. Kim and J.~C. Kim, ``Vibrotactile rendering for a traveling vibrotactile
  wave based on a haptic processor,'' \emph{IEEE Transactions on Haptics},
  vol.~5, no.~1, pp. 14--20, 2012.

\bibitem{Seo2010}
J.~Seo and S.~Choi, ``Initial study of creating linearly moving vibrotactile
  sensation on mobile device,'' in \emph{Proceedings of the Haptics Symposium
  (HS)}.\hskip 1em plus 0.5em minus 0.4em\relax IEEE, 2010, pp. 67--70.

\bibitem{Seo2013}
------, ``Perceptual analysis of vibrotactile flows on a mobile device,''
  \emph{IEEE Transactions on Haptics}, vol.~6, no.~4, pp. 522--527, 2013.

\bibitem{KangKookChoEtAl2012}
J.~Kang, J.~Kook, K.~Cho, S.~Wang, and J.~Ryu, ``Effects of amplitude
  modulation on vibrotactile flow displays on piezo-actuated thin touch
  screen,'' \emph{International Journal of Control, Automation, and Systems},
  vol.~10, no.~3, pp. 582--588, 2012.

\bibitem{Kang2012}
J.~Kang, J.~Lee, H.~Kim, K.~Cho, S.~Wang, and J.~Ryu, ``Smooth vibrotactile
  flow generation using two piezoelectric actuators,'' \emph{IEEE Transactions
  on Haptics}, vol.~5, no.~1, pp. 21--32, 2012.

\bibitem{Seo2015_WHC}
J.~Seo and S.~Choi, ``Edge flows: Improving information transmission in mobile
  devices using two-dimensional vibrotactile flows,'' in \emph{Proceedings of
  the IEEE World Haptics Conference (WHC)}, 2015, pp. 25--30.

\bibitem{nara:2000}
T.~{Nara}, M.~{Takasaki}, S.~{Tachi}, and T.~{Higuchi}, ``An application of saw
  to a tactile display in virtual reality,'' in \emph{2000 IEEE Ultrasonics
  Symposium. Proceedings. An International Symposium}, vol.~1, Oct 2000, pp.
  1--4 vol.1.

\bibitem{gray1875improvement}
E.~Gray, ``Improvement in electric telegraphs for transmitting musical,
  tones,'' Jul. 1875, uS Patent US166,096.

\bibitem{Johnsen23}
A.~{Johnsen} and K.~{Rahbek}, ``A physical phenomenon and its applications to
  telegraphy, telephony, etc.'' \emph{Journal of the Institution of Electrical
  Engineers}, vol.~61, no. 320, pp. 713--725, July 1923.

\bibitem{Mallinckrodt53}
E.~Mallinckrodt, A.~L. Hughes, and W.~Sleator, ``Perception by the skin of
  electrically induced vibrations,'' \emph{Science}, vol. 118, no. 3062, pp.
  277--278, 1953.

\bibitem{tang1998microfabricated}
H.~Tang and D.~J. Beebe, ``A microfabricated electrostatic haptic display for
  persons with visual impairments,'' \emph{IEEE Transactions on Rehabilitation
  Engineering}, vol.~6, no.~3, pp. 241--248, 1998.

\bibitem{Bau10}
O.~Bau, I.~Poupyrev, A.~Israr, and C.~Harrison, ``Teslatouch: Electrovibration
  for touch surfaces,'' in \emph{Proceedings of the 23Nd Annual ACM Symposium
  on User Interface Software and Technology}, ser. UIST '10, 2010, pp.
  283--292.

\bibitem{linjama2009sense}
J.~Linjama and V.~M{\"a}kinen, ``E-sense screen: Novel haptic display with
  capacitive electrosensory interface,'' in \emph{HAID 2009, 4th Workshop for
  Haptic and Audio Interaction Design}, 2009.

\bibitem{Bau12}
O.~Bau and I.~Poupyrev, ``Revel: Tactile feedback technology for augmented
  reality,'' \emph{ACM Trans. Graph.}, vol.~31, no.~4, pp. 89:1--89:11, Jul.
  2012.

\bibitem{Nakamura16}
T.~{Nakamura} and A.~{Yamamoto}, ``A multi-user surface visuo-haptic display
  using electrostatic friction modulation and capacitive-type position
  sensing,'' \emph{IEEE Transactions on Haptics}, vol.~9, no.~3, pp. 311--322,
  July 2016.

\bibitem{kim2015method}
H.~Kim, J.~Kang, K.-D. Kim, K.-M. Lim, and J.~Ryu, ``Method for providing
  electrovibration with uniform intensity,'' \emph{IEEE Transactions on
  Haptics}, vol.~8, no.~4, pp. 492--496, 2015.

\bibitem{Marchuk:2010}
N.~D. {Marchuk}, J.~E. {Colgate}, and M.~A. {Peshkin}, ``Friction measurements
  on a large area tpad,'' in \emph{2010 IEEE Haptics Symposium}, March 2010,
  pp. 317--320.

\bibitem{giraud:2012}
F.~{Giraud}, M.~{Amberg}, B.~{Lemaire-Semail}, and G.~{casiez}, ``Design of a
  transparent tactile stimulator,'' in \emph{2012 IEEE Haptics Symposium
  (HAPTICS)}, March 2012, pp. 485--489.

\bibitem{casset:2013}
F.~{Casset}, J.~S. {Danel}, C.~{Chappaz}, Y.~{Civet}, M.~{Amberg},
  M.~{Gorisse}, C.~{Dieppedale}, G.~{Le Rhun}, S.~{Basrour}, P.~{Renaux},
  E.~{Defaÿ}, A.~{Devos}, B.~{Semail}, P.~{Ancey}, and S.~{Fanget}, ``Low
  voltage actuated plate for haptic applications with pzt thin-film,'' in
  \emph{2013 Transducers Eurosensors XXVII: The 17th International Conference
  on Solid-State Sensors, Actuators and Microsystems (TRANSDUCERS EUROSENSORS
  XXVII)}, June 2013, pp. 2733--2736.

\bibitem{Mullenbach:2013}
J.~Mullenbach, C.~Shultz, A.~M. Piper, M.~Peshkin, and J.~E. Colgate, ``Surface
  haptic interactions with a tpad tablet,'' in \emph{Proceedings of the Adjunct
  Publication of the 26th Annual ACM Symposium on User Interface Software and
  Technology}, ser. UIST '13 Adjunct, 2013, pp. 7--8.

\bibitem{fujitsu}
``Fujitsu journal,'' https://www.fujitsu.com/global/about/\\
  resources/news/press-releases/2014/0224-01.html, accessed on November 1st
  2019.

\bibitem{Yang:2015}
{Yang, Yi}, {Lemaire-Semail, Betty}, {Giraud, Fr\'ed\'eric}, {Amberg, Michel},
  {Zhang, Yuru}, and {Giraud-Audine, Christophe}, ``Power analysis for the
  design of a large area ultrasonic tactile touch panel,'' \emph{Eur. Phys. J.
  Appl. Phys.}, vol.~72, p.~12, 2015.

\bibitem{vezzoli:2016}
E.~Vezzoli, T.~Sednaoui, M.~Amberg, F.~Giraud, and B.~Lemaire-Semail, ``Texture
  rendering strategies with a high fidelity - capacitive visual-haptic friction
  control device,'' in \emph{Haptics: Perception, Devices, Control, and
  Applications}, F.~Bello, H.~Kajimoto, and Y.~Visell, Eds.\hskip 1em plus
  0.5em minus 0.4em\relax Springer International Publishing, 2016, pp.
  251--260.

\bibitem{giraud:2018}
F.~{Giraud}, T.~{Hara}, C.~{Giraud-Audine}, M.~{Amberg}, B.~{Lemaire-Semail},
  and M.~{Takasaki}, ``Evaluation of a friction reduction based haptic surface
  at high frequency,'' in \emph{2018 IEEE Haptics Symposium (HAPTICS)}, March
  2018, pp. 210--215.

\bibitem{sirin2019fingerpad}
O.~Sirin, A.~Barrea, P.~Lef{\`e}vre, J.-L. Thonnard, and C.~Basdogan,
  ``Fingerpad contact evolution under electrovibration,'' \emph{Journal of the
  Royal Society Interface}, vol.~16, no. 156, p. 20190166, 2019.

\bibitem{tomlinson2009understanding}
S.~E. Tomlinson, ``Understanding the friction between human fingers and
  contacting surfaces,'' Ph.D. dissertation, University of Sheffield, 2009.

\bibitem{adams2013finger}
M.~J. Adams, S.~A. Johnson, P.~Lef{\`e}vre, V.~L{\'e}vesque, V.~Hayward,
  T.~Andr{\'e}, and J.-L. Thonnard, ``Finger pad friction and its role in grip
  and touch,'' \emph{Journal of The Royal Society Interface}, vol.~10, no.~80,
  p. 20120467, 2013.

\bibitem{van2015review}
J.~van Kuilenburg, M.~A. Masen, and E.~van~der Heide, ``A review of fingerpad
  contact mechanics and friction and how this affects tactile perception,''
  \emph{Proceedings of the Institution of Mechanical Engineers, Part J: Journal
  of engineering tribology}, vol. 229, no.~3, pp. 243--258, 2015.

\bibitem{sahli2018evolution}
R.~Sahli, G.~Pallares, C.~Ducottet, I.~B. Ali, S.~Al~Akhrass, M.~Guibert, and
  J.~Scheibert, ``Evolution of real contact area under shear and the value of
  static friction of soft materials,'' \emph{Proceedings of the National
  Academy of Sciences}, vol. 115, no.~3, pp. 471--476, 2018.

\bibitem{vodlak2016multi}
T.~Vodlak, Z.~Vidrih, E.~Vezzoli, B.~Lemaire-Semail, and D.~Peric,
  ``Multi-physics modelling and experimental validation of electrovibration
  based haptic devices,'' \emph{Biotribology}, vol.~8, pp. 12--25, 2016.

\bibitem{bowden2001friction}
F.~P. Bowden, F.~P. Bowden, and D.~Tabor, \emph{The Friction and Lubrication of
  Solids}.\hskip 1em plus 0.5em minus 0.4em\relax Oxford university press,
  2001, vol.~1.

\bibitem{Adams07}
M.~J. Adams, B.~J. Briscoe, and S.~A. Johnson, ``Friction and lubrication of
  human skin,'' \emph{Tribology Letters}, vol.~26, no.~3, pp. 239--253, Jun
  2007.

\bibitem{bochereau2017characterizing}
S.~Bochereau, B.~Dzidek, M.~Adams, and V.~Hayward, ``Characterizing and imaging
  gross and real finger contacts under dynamic loading,'' \emph{IEEE
  Transactions on Haptics}, vol.~10, no.~4, pp. 456--465, 2017.

\bibitem{persson2006contact}
B.~N. Persson, ``Contact mechanics for randomly rough surfaces,'' \emph{Surface
  Science Reports}, vol.~61, no.~4, pp. 201--227, 2006.

\bibitem{persson2018dependency}
B.~Persson, ``The dependency of adhesion and friction on electrostatic
  attraction,'' \emph{The Journal of Chemical Physics}, vol. 148, no.~14, p.
  144701, 2018.

\bibitem{ayyildiz2018contact}
M.~Ayyildiz, M.~Scaraggi, O.~Sirin, C.~Basdogan, and B.~N. Persson, ``Contact
  mechanics between the human finger and a touchscreen under electroadhesion,''
  \emph{Proceedings of the National Academy of Sciences}, vol. 115, no.~50, pp.
  12\,668--12\,673, 2018.

\bibitem{sirin2019electroadhesion}
O.~Sirin, M.~Ayyildiz, B.~Persson, and C.~Basdogan, ``Electroadhesion with
  application to touchscreens,'' \emph{Soft Matter}, vol.~15, no.~8, pp.
  1758--1775, 2019.

\bibitem{shultz2018electrical}
C.~D. Shultz, M.~A. Peshkin, and J.~E. Colgate, ``On the electrical
  characterization of electroadhesive displays and the prominent interfacial
  gap impedance associated with sliding fingertips,'' in \emph{2018 IEEE
  Haptics Symposium (HAPTICS)}, 2018, pp. 151--157.

\bibitem{Watanabe95}
T.~{Watanabe} and S.~{Fukui}, ``A method for controlling tactile sensation of
  surface roughness using ultrasonic vibration,'' in \emph{Proceedings of 1995
  IEEE International Conference on Robotics and Automation}, vol.~1, May 1995,
  pp. 1134--1139 vol.1.

\bibitem{Vezzoli:2017}
E.~{Vezzoli}, Z.~{Vidrih}, V.~{Giamundo}, B.~{Lemaire-Semail}, F.~{Giraud},
  T.~{Rodic}, D.~{Peric}, and M.~{Adams}, ``Friction reduction through
  ultrasonic vibration part 1: Modelling intermittent contact,'' \emph{IEEE
  Transactions on Haptics}, vol.~10, no.~2, pp. 196--207, April 2017.

\bibitem{Wiertlewski16}
M.~Wiertlewski, R.~F. Friesen, and J.~E. Colgate, ``Partial squeeze film
  levitation modulates fingertip friction,'' \emph{Proceedings of the National
  Academy of Sciences}, vol. 113, no.~33, pp. 9210--9215, 2016.

\bibitem{giraud:2010}
F.~Giraud, M.~Amberg, R.~Vanbelleghem, and B.~Lemaire-Semail, ``Power
  consumption reduction of a controlled friction tactile plate,'' in
  \emph{Haptics: Generating and Perceiving Tangible Sensations}, A.~M.~L.
  Kappers, J.~B.~F. van Erp, W.~M. Bergmann~Tiest, and F.~C.~T. van~der Helm,
  Eds.\hskip 1em plus 0.5em minus 0.4em\relax Berlin, Heidelberg: Springer
  Berlin Heidelberg, 2010, pp. 44--49.

\bibitem{Wiertlewski15}
M.~{Wiertlewski} and J.~E. {Colgate}, ``Power optimization of ultrasonic
  friction-modulation tactile interfaces,'' \emph{IEEE Transactions on
  Haptics}, vol.~8, no.~1, pp. 43--53, Jan 2015.

\bibitem{Vardar17}
Y.~{Vardar}, B.~{Güçlü}, and C.~{Basdogan}, ``Effect of waveform on tactile
  perception by electrovibration displayed on touch screens,'' \emph{IEEE
  Transactions on Haptics}, vol.~10, no.~4, pp. 488--499, Oct 2017.

\bibitem{wiert:2014}
M.~Wiertlewski, D.~Leonardis, D.~J. Meyer, M.~A. Peshkin, and J.~E. Colgate,
  ``A high-fidelity surface-haptic device for texture rendering on bare
  finger,'' in \emph{Haptics: Neuroscience, Devices, Modeling, and
  Applications}, M.~Auvray and C.~Duriez, Eds.\hskip 1em plus 0.5em minus
  0.4em\relax Berlin, Heidelberg: Springer Berlin Heidelberg, 2014, pp.
  241--248.

\bibitem{messaoud:2016b}
W.~{Ben Messaoud}, F.~{Giraud}, B.~{Lemaire-Semail}, M.~{Amberg}, and
  M.~{Bueno}, ``Amplitude control of an ultrasonic vibration for a tactile
  stimulator,'' \emph{IEEE/ASME Transactions on Mechatronics}, vol.~21, no.~3,
  pp. 1692--1701, June 2016.

\bibitem{Mullenbach:2016}
J.~{Mullenbach}, M.~{Peshkin}, and J.~E. {Colgate}, ``eshiver: Lateral force
  feedback on fingertips through oscillatory motion of an electroadhesive
  surface,'' \emph{IEEE Transactions on Haptics}, vol.~10, no.~3, pp. 358--370,
  2017.

\bibitem{Jeon2009}
S.~Jeon and S.~Choi, ``Haptic augmented reality: Taxonomy and an example of
  stiffness modulation,'' \emph{Presence: Teleoperators and Virtual
  Environments}, vol.~18, no.~5, pp. 387--408, 2009.

\bibitem{Bolanowski88}
S.~J. Bolanowski, Jr., G.~A. Gesheider, R.~T. Verrillo, and C.~M. Checkosky,
  ``Four channels mediate the mechanical aspects of touch,'' \emph{Journal of
  Acoustical Society of America}, vol.~84, no.~5, pp. 1680--1694, 1988.

\bibitem{Gescheider2008}
G.~A. Gescheider, J.~H. Wright, and R.~T. Verrillo,
  \emph{Information-Processing Channels in the Tactile Sensory System: A
  Psychophysical and Physiological Analysis}.\hskip 1em plus 0.5em minus
  0.4em\relax Psychology Press, 2008.

\bibitem{Jones2008}
L.~A. Jones and N.~B. Sarter, ``Tactile displays: Guidance for their design and
  application,'' \emph{Human Factors}, vol.~50, no.~1, pp. 90--111, 2008.

\bibitem{Johansson2009}
R.~S. Johansson and J.~R. Flanagan, ``Coding and use of tactile signals from
  theffingertips in object manipulation tasks,'' \emph{Nature Reviews
  Neuroscience}, vol.~10, pp. 345--359, 2009.

\bibitem{Brisben99}
A.~J. Brisben, S.~S. Hsiao, and K.~O. Johnson, ``Detection of vibration
  transmitted through an object grasped in the hand,'' \emph{Journal of
  Neurophysiology}, vol.~81, no.~4, pp. 1548--1558, 1999.

\bibitem{Derler09}
S.~Derler, L.-C. Gerhardt, A.~Lenz, E.~Bertaux, and M.~Hadad, ``Friction of
  human skin against smooth and rough glass as a function of the contact
  pressure,'' \emph{Tribology International}, vol.~42, no.~11, pp. 1565 --
  1574, 2009, special Issue: 35th Leeds-Lyon Symposium.

\bibitem{derler2012tribology}
S.~Derler and L.-C. Gerhardt, ``Tribology of skin: review and analysis of
  experimental results for the friction coefficient of human skin,''
  \emph{Tribology Letters}, vol.~45, no.~1, pp. 1--27, 2012.

\bibitem{wijekoon2012electrostatic}
D.~Wijekoon, M.~E. Cecchinato, E.~Hoggan, and J.~Linjama, ``Electrostatic
  modulated friction as tactile feedback: intensity perception,'' in
  \emph{International Conference on Human Haptic Sensing and Touch Enabled
  Computer Applications}, 2012, pp. 613--624.

\bibitem{Vardar16}
Y.~Vardar, B.~G{\"u}{\c{c}}l{\"u}, and C.~Basdogan, ``Effect of waveform in
  haptic perception of electrovibration on touchscreens,'' in \emph{Haptics:
  Perception, Devices, Control, and Applications}, F.~Bello, H.~Kajimoto, and
  Y.~Visell, Eds.\hskip 1em plus 0.5em minus 0.4em\relax Cham: Springer
  International Publishing, 2016, pp. 190--203.

\bibitem{Vardar18}
Y.~{Vardar}, B.~{Güçlü}, and C.~{Basdogan}, ``Tactile masking by
  electrovibration,'' \emph{IEEE Transactions on Haptics}, vol.~11, no.~4, pp.
  623--635, Oct 2018.

\bibitem{ryu2018mechanical}
S.~Ryu, D.~Pyo, S.-C. Lim, and D.-S. Kwon, ``Mechanical vibration influences
  the perception of electrovibration,'' \emph{Scientific reports}, vol.~8,
  no.~1, pp. 1--10, 2018.

\bibitem{jamalzadeh2019effect}
M.~Jamalzadeh, B.~G{\"u}{\c{c}}l{\"u}, Y.~Vardar, and C.~Basdogan, ``Effect of
  remote masking on detection of electrovibration,'' in \emph{2019 IEEE World
  Haptics Conference (WHC)}, 2019, pp. 229--234.

\bibitem{vardar2017roughness}
Y.~Vardar, A.~{\.I}{\c{s}}leyen, M.~K. Saleem, and C.~Basdogan, ``Roughness
  perception of virtual textures displayed by electrovibration on touch
  screens,'' in \emph{2017 IEEE World Haptics Conference (WHC)}, 2017, pp.
  263--268.

\bibitem{smith2002role}
A.~M. Smith, C.~E. Chapman, M.~Deslandes, J.-S. Langlais, and M.-P. Thibodeau,
  ``Role of friction and tangential force variation in the subjective scaling
  of tactile roughness,'' \emph{Experimental brain research}, vol. 144, no.~2,
  pp. 211--223, 2002.

\bibitem{Isleyen19}
\BIBentryALTinterwordspacing
A.~{\.I}{\c{s}}leyen, Y.~{Vardar}, and C.~{Basdogan}, ``Tactile roughness
  perception of virtual gratings by electrovibration,'' \emph{IEEE Transactions
  on Haptics}, 2019. [Online]. Available:
  \url{http://ieeexplore.ieee.org/document/8933496}
\BIBentrySTDinterwordspacing

\bibitem{ito2019tactile}
K.~Ito, S.~Okamoto, Y.~Yamada, and H.~Kajimoto, ``Tactile texture display with
  vibrotactile and electrostatic friction stimuli mixed at appropriate ratio
  presents better roughness textures,'' \emph{ACM Transactions on Applied
  Perception (TAP)}, vol.~16, no.~4, pp. 1--15, 2019.

\bibitem{ozdamar2020step}
I.~Ozdamar, M.~R. Alipour, B.~P. Delhaye, P.~Lefevre, and C.~Basdogan,
  ``Step-change in friction under electrovibration,'' \emph{IEEE Transactions
  on Haptics}, vol.~13, no.~1, pp. 137--143, 2020.

\bibitem{Mun2019}
S.~Mun, H.~Lee, and S.~Choi, ``Perceptual space of regular homogeneous haptic
  textures rendered using electrovibration,'' in \emph{Proceedings of the IEEE
  World Haptics Conference}, 2019, pp. 7--12.

\bibitem{samur2009psychophysical}
E.~Samur, J.~E. Colgate, and M.~A. Peshkin, ``Psychophysical evaluation of a
  variable friction tactile interface,'' in \emph{Human Vision and Electronic
  Imaging XIV}, vol. 7240.\hskip 1em plus 0.5em minus 0.4em\relax International
  Society for Optics and Photonics, 2009, p. 72400J.

\bibitem{messaoud2016relation}
W.~B. Messaoud, M.-A. Bueno, and B.~Lemaire-Semail, ``Relation between human
  perceived friction and finger friction characteristics,'' \emph{Tribology
  International}, vol.~98, pp. 261--269, 2016.

\bibitem{saleem2017tactile}
M.~K. Saleem, C.~Yilmaz, and C.~Basdogan, ``Tactile perception of change in
  friction on an ultrasonically actuated glass surface,'' in \emph{2017 IEEE
  World Haptics Conference (WHC)}, 2017, pp. 495--500.

\bibitem{Saleem18}
M.~K. {Saleem}, C.~{Yilmaz}, and C.~{Basdogan}, ``Psychophysical evaluation of
  change in friction on an ultrasonically-actuated touchscreen,'' \emph{IEEE
  Transactions on Haptics}, vol.~11, no.~4, pp. 599--610, Oct 2018.

\bibitem{gueorguiev2017feeling}
D.~Gueorguiev, E.~Vezzoli, T.~Sednaoui, L.~Grisoni, and B.~Lemaire-Semail,
  ``Feeling multiple edges: the tactile perception of short ultrasonic square
  reductions of the finger-surface friction,'' in \emph{2017 IEEE World Haptics
  Conference (WHC)}, 2017, pp. 125--129.

\bibitem{gueorguiev2017tactile}
D.~Gueorguiev, E.~Vezzoli, A.~Mouraux, B.~Lemaire-Semail, and J.-L. Thonnard,
  ``The tactile perception of transient changes in friction,'' \emph{Journal of
  The Royal Society Interface}, vol.~14, no. 137, p. 20170641, 2017.

\bibitem{biet2008discrimination}
M.~Biet, G.~Casiez, F.~Giraud, and B.~Lemaire-Semail, ``Discrimination of
  virtual square gratings by dynamic touch on friction based tactile
  displays,'' in \emph{Symposium on Haptic Interfaces for Virtual Environment
  and Teleoperator Systems}, 2008, pp. 41--48.

\bibitem{saleemtactile}
\BIBentryALTinterwordspacing
M.~K. Saleem, C.~Yilmaz, and C.~Basdogan, ``Tactile perception of virtual edges
  and gratings displayed by friction modulation via ultrasonic actuation,''
  \emph{IEEE Transactions on Haptics}, 2019. [Online]. Available:
  \url{https://ieeexplore.ieee.org/document/8883061}
\BIBentrySTDinterwordspacing

\bibitem{US9196134}
V.~Lévesque and J.~M. Cruz-Hernandez, ``Method and apparatus for simulating
  surface features on a user interface with haptic effects,'' U.S. Patent
  9,196,134, Nov. 24, 2015.

\bibitem{Jiao18}
J.~{Jiao}, Y.~{Zhang}, D.~{Wang}, Y.~{Visell}, D.~{Cao}, X.~{Guo}, and
  X.~{Sun}, ``Data-driven rendering of fabric textures on electrostatic tactile
  displays,'' in \emph{2018 IEEE Haptics Symposium (HAPTICS)}, March 2018, pp.
  169--174.

\bibitem{osgouei2018inverse}
R.~H. Osgouei, S.~Shin, J.~R. Kim, and S.~Choi, ``An inverse neural network
  model for data-driven texture rendering on electrovibration display,'' in
  \emph{2018 IEEE Haptics Symposium (HAPTICS)}, 2018, pp. 270--277.

\bibitem{Osgouei2020}
\BIBentryALTinterwordspacing
R.~H. Osgouei, J.~R. Kim, and S.~Choi, ``Data-driven texture modeling and
  rendering on electrovibration display,'' \emph{IEEE Transactions on Haptics},
  2020. [Online]. Available: \url{https://ieeexplore.ieee.org/document/8787895}
\BIBentrySTDinterwordspacing

\bibitem{Messaoud16}
W.~B. Messaoud, M.-A. Bueno, and B.~Lemaire-Semail, ``Textile fabrics' texture:
  From multi-level feature extraction to tactile simulation,'' in
  \emph{Haptics: Perception, Devices, Control, and Applications}, F.~Bello,
  H.~Kajimoto, and Y.~Visell, Eds.\hskip 1em plus 0.5em minus 0.4em\relax Cham:
  Springer International Publishing, 2016, pp. 294--303.

\bibitem{Camillieri18}
B.~Camillieri, M.-A. Bueno, M.~Fabre, B.~Juan, B.~Lemaire-Semail, and
  L.~Mouchnino, ``From finger friction and induced vibrations to brain
  activation: Tactile comparison between real and virtual textile fabrics,''
  \emph{Tribology International}, vol. 126, pp. 283 -- 296, 2018.

\bibitem{wu2017tactile}
S.~Wu, X.~Sun, Q.~Wang, and J.~Chen, ``Tactile modeling and rendering
  image-textures based on electrovibration,'' \emph{The Visual Computer},
  vol.~33, no.~5, pp. 637--646, 2017.

\bibitem{Xu11}
C.~Xu, A.~Israr, I.~Poupyrev, O.~Bau, and C.~Harrison, ``Tactile display for
  the visually impaired using teslatouch,'' in \emph{CHI '11 Extended Abstracts
  on Human Factors in Computing Systems}, ser. CHI EA '11, 2011, pp. 317--322.

\bibitem{Kim13}
S.-C. Kim, A.~Israr, and I.~Poupyrev, ``Tactile rendering of 3d features on
  touch surfaces,'' in \emph{Proceedings of the 26th Annual ACM Symposium on
  User Interface Software and Technology}, ser. UIST '13, 2013, pp. 531--538.

\bibitem{Osgouei17}
R.~H. {Osgouei}, J.~R. {Kim}, and S.~{Choi}, ``Improving 3d shape recognition
  with electrostatic friction display,'' \emph{IEEE Transactions on Haptics},
  vol.~10, no.~4, pp. 533--544, Oct 2017.

\bibitem{Ware14}
J.~{Ware}, E.~{Cha}, M.~A. {Peshkin}, J.~E. {Colgate}, and R.~L. {Klatzky},
  ``Search efficiency for tactile features rendered by surface haptic
  displays,'' \emph{IEEE Transactions on Haptics}, vol.~7, no.~4, pp. 545--550,
  2014.

\bibitem{moussette2012simple}
C.~Moussette, ``Simple haptics: Sketching perspectives for the design of haptic
  interactions,'' Ph.D. dissertation, Ume{\aa} Universitet, 2012.

\bibitem{schneider2017haptic}
O.~Schneider, K.~MacLean, C.~Swindells, and K.~Booth, ``Haptic experience
  design: What hapticians do and where they need help,'' \emph{International
  Journal of Human-Computer Studies}, vol. 107, pp. 5--21, 2017.

\bibitem{buxton2010sketching}
B.~Buxton, \emph{Sketching User Experiences: Getting the Design Right and the
  Right Design}.\hskip 1em plus 0.5em minus 0.4em\relax Morgan Kaufmann, 2010.

\bibitem{Potier16}
\BIBentryALTinterwordspacing
L.~Potier, T.~Pietrzak, G.~Casiez, and N.~Roussel, ``Designing tactile patterns
  with programmable friction,'' in \emph{Actes De La 28I\`{e}Me Conference
  Francophone Sur L'Interaction Homme-Machine}, ser. IHM '16.\hskip 1em plus
  0.5em minus 0.4em\relax New York, NY, USA: ACM, 2016, pp. 1--7. [Online].
  Available: \url{http://doi.acm.org/10.1145/3004107.3004110}
\BIBentrySTDinterwordspacing

\bibitem{Levesque11}
V.~Levesque, L.~Oram, K.~MacLean, A.~Cockburn, N.~D. Marchuk, D.~Johnson, J.~E.
  Colgate, and M.~A. Peshkin, ``Enhancing physicality in touch interaction with
  programmable friction,'' in \emph{Proceedings of the SIGCHI Conference on
  Human Factors in Computing Systems}, ser. CHI '11, 2011, pp. 2481--2490.

\bibitem{maclean17lessons}
K.~MacLean, ``Special session on lessons learned,'' IEEE World Haptics
  Conference, 2017.

\bibitem{Mullenbach14}
J.~Mullenbach, C.~Shultz, J.~E. Colgate, and A.~M. Piper, ``Exploring affective
  communication through variable-friction surface haptics,'' in
  \emph{Proceedings of the SIGCHI Conference on Human Factors in Computing
  Systems}, ser. CHI '14, 2014, pp. 3963--3972.

\bibitem{Zhang15}
Y.~Zhang and C.~Harrison, ``Quantifying the targeting performance benefit of
  electrostatic haptic feedback on touchscreens,'' in \emph{Proceedings of the
  2015 International Conference on Interactive Tabletops \& Surfaces}, ser. ITS
  '15, 2015, pp. 43--46.

\bibitem{Casiez11}
G.~Casiez, N.~Roussel, R.~Vanbelleghem, and F.~Giraud, ``Surfpad: Riding
  towards targets on a squeeze film effect,'' in \emph{Proceedings of the
  SIGCHI Conference on Human Factors in Computing Systems}, ser. CHI '11, 2011,
  pp. 2491--2500.

\bibitem{nashel2003tactile}
A.~Nashel and S.~Razzaque, ``Tactile virtual buttons for mobile devices,'' in
  \emph{CHI'03 Extended Abstracts on Human Factors in Computing Systems}, 2003,
  pp. 854--855.

\bibitem{brewster2007tactile}
S.~Brewster, S.~Brewster, F.~Chohan, and L.~Brown, ``Tactile feedback for
  mobile interactions,'' in \emph{Proceedings of the SIGCHI Conference on Human
  Factors in Computing Systems}, 2007, pp. 159--162.

\bibitem{hoggan2008crossmodal}
E.~Hoggan, T.~Kaaresoja, P.~Laitinen, and S.~Brewster, ``Crossmodal congruence:
  the look, feel and sound of touchscreen widgets,'' in \emph{Proceedings of
  the 10th International Conference on Multimodal Interfaces}, 2008, pp.
  157--164.

\bibitem{park2011tactile}
G.~Park, S.~Choi, K.~Hwang, S.~Kim, J.~Sa, and M.~Joung, ``Tactile effect
  design and evaluation for virtual buttons on a mobile device touchscreen,''
  in \emph{Proceedings of the 13th International Conference on Human Computer
  Interaction with Mobile Devices and Services}, 2011, pp. 11--20.

\bibitem{pakkanen2010comparison}
T.~Pakkanen, R.~Raisamo, J.~Raisamo, K.~Salminen, and V.~Surakka, ``Comparison
  of three designs for haptic button edges on touchscreens,'' in \emph{2010
  IEEE Haptics Symposium}, 2010, pp. 219--225.

\bibitem{chen2011design}
H.-Y. Chen, J.~Park, S.~Dai, and H.~Z. Tan, ``Design and evaluation of
  identifiable key-click signals for mobile devices,'' \emph{IEEE Transactions
  on Haptics}, vol.~4, no.~4, pp. 229--241, 2011.

\bibitem{lylykangas2011designing}
J.~Lylykangas, V.~Surakka, K.~Salminen, J.~Raisamo, P.~Laitinen,
  K.~R{\"o}nning, and R.~Raisamo, ``Designing tactile feedback for piezo
  buttons,'' in \emph{Proceedings of the SIGCHI Conference on Human Factors in
  Computing Systems}, 2011, pp. 3281--3284.

\bibitem{kim2014study}
J.~R. Kim and H.~Z. Tan, ``A study of touch typing performance with keyclick
  feedback,'' in \emph{2014 IEEE Haptics Symposium (HAPTICS)}, 2014, pp.
  227--233.

\bibitem{wu2015haptic}
C.-M. Wu and S.~Smith, ``A haptic keypad design with a novel interactive haptic
  feedback method,'' \emph{Journal of Engineering Design}, vol.~26, no. 4-6,
  pp. 169--186, 2015.

\bibitem{ma2015haptic}
Z.~Ma, D.~Edge, L.~Findlater, and H.~Z. Tan, ``Haptic keyclick feedback
  improves typing speed and reduces typing errors on a flat keyboard,'' in
  \emph{2015 IEEE World Haptics Conference (WHC)}, 2015, pp. 220--227.

\bibitem{Sadia19}
\BIBentryALTinterwordspacing
B.~Sadia, S.~E. Emgin, T.~M. Sezgin, and C.~Basdogan, ``Data-driven
  vibrotactile rendering of digital buttons on touchscreens,''
  \emph{International Journal of Human-Computer Studies}, 2019. [Online].
  Available: \url{https://doi.org/10.1016/j.ijhcs.2019.09.005}
\BIBentrySTDinterwordspacing

\bibitem{Banter:2010}
B.~Banter, ``Touch screens and touch surfaces are enriched by haptic
  force-feedback,'' \emph{Information Display}, vol.~26, no.~3, pp. 26--30,
  2010.

\bibitem{tashiro2009realization}
K.~Tashiro, Y.~Shiokawa, T.~Aono, and T.~Maeno, ``Realization of button click
  feeling by use of ultrasonic vibration and force feedback,'' in \emph{World
  Haptics 2009-Third Joint EuroHaptics conference and Symposium on Haptic
  Interfaces for Virtual Environment and Teleoperator Systems}, 2009, pp. 1--6.

\bibitem{monnoyer2017optimal}
J.~Monnoyer, E.~Diaz, C.~Bourdin, and M.~Wiertlewski, ``Optimal skin impedance
  promotes perception of ultrasonic switches,'' in \emph{2017 IEEE World
  Haptics Conference (WHC)}, 2017, pp. 130--135.

\bibitem{monnoyer2016ultrasonic}
------, ``Ultrasonic friction modulation while pressing induces a tactile
  feedback,'' in \emph{International Conference on Human Haptic Sensing and
  Touch Enabled Computer Applications}, 2016, pp. 171--179.

\bibitem{gueorguiev:2018}
D.~Gueorguiev, A.~Kaci, M.~Amberg, F.~Giraud, and B.~Lemaire-Semail,
  ``Travelling ultrasonic wave enhances keyclick sensation,'' in \emph{Haptics:
  Science, Technology, and Applications}, D.~Prattichizzo, H.~Shinoda, H.~Z.
  Tan, E.~Ruffaldi, and A.~Frisoli, Eds.\hskip 1em plus 0.5em minus 0.4em\relax
  Springer International Publishing, 2018.

\bibitem{Levesque12}
V.~{Lévesque}, L.~{Oram}, and K.~{MacLean}, ``Exploring the design space of
  programmable friction for scrolling interactions,'' in \emph{2012 IEEE
  Haptics Symposium (HAPTICS)}, March 2012, pp. 23--30.

\bibitem{Giraud13a}
F.~Giraud, M.~Amberg, and B.~Lemaire-Semail, ``Design and control of a haptic
  knob,'' \emph{Sensors and Actuators A: Physical}, vol. 196, pp. 78 -- 85,
  2013.

\bibitem{US9330544}
V.~Lévesque, J.~M. Cruz-Hernandez, A.~Weddle, and D.~Birnbaum, ``System and
  method for simulated physical interactions with haptic effects,'' U.S. Patent
  9,330,544, March 3, 2013.

\bibitem{soukoreff2004towards}
R.~W. Soukoreff and I.~S. MacKenzie, ``Towards a standard for pointing device
  evaluation, perspectives on 27 years of fitts’ law research in hci,''
  \emph{International Journal of Human-Computer Studies}, vol.~61, no.~6, pp.
  751--789, 2004.

\bibitem{Tory13}
M.~Tory and R.~Kincaid, ``Comparing physical, overlay, and touch screen
  parameter controls,'' in \emph{Proceedings of the 2013 ACM International
  Conference on Interactive Tabletops and Surfaces}.\hskip 1em plus 0.5em minus
  0.4em\relax ACM, 2013, pp. 91--100.

\bibitem{hilliges2007photohelix}
O.~Hilliges, D.~Baur, and A.~Butz, ``Photohelix: Browsing, sorting and sharing
  digital photo collections,'' in \emph{Second Annual IEEE International
  Workshop on Horizontal Interactive Human-Computer Systems (TABLETOP'07)},
  2007, pp. 87--94.

\bibitem{weiss2009slap}
M.~Weiss, R.~Jennings, R.~Khoshabeh, J.~Borchers, J.~Wagner, Y.~Jansen, and
  J.~D. Hollan, ``Slap widgets: bridging the gap between virtual and physical
  controls on tabletops,'' in \emph{CHI'09 Extended Abstracts on Human Factors
  in Computing Systems}.\hskip 1em plus 0.5em minus 0.4em\relax ACM, 2009, pp.
  3229--3234.

\bibitem{Israr12}
A.~Israr, O.~Bau, S.-C. Kim, and I.~Poupyrev, ``Tactile feedback on flat
  surfaces for the visually impaired,'' in \emph{CHI '12 Extended Abstracts on
  Human Factors in Computing Systems}, ser. CHI EA '12, 2012, pp. 1571--1576.

\bibitem{Lim2019}
J.~Lim, Y.~Yoo, H.~Cho, and S.~Choi, ``Touchphoto: Enabling independent picture
  taking and understanding for visually-impaired users,'' in \emph{Proceedings
  of the ACM International Conference on Multimodal Interaction}, 2019, pp.
  124--134.

\bibitem{guo2019effect}
\BIBentryALTinterwordspacing
X.~Guo, Y.~Zhang, W.~Wei, W.~Xu, and D.~Wang, ``Effect of temperature on the
  absolute and discrimination thresholds of voltage on electrovibration tactile
  display,'' \emph{IEEE Transactions on Haptics}, 2019. [Online]. Available:
  \url{https://ieeexplore.ieee.org/document/8944057}
\BIBentrySTDinterwordspacing

\bibitem{vezzoli2015physical}
E.~Vezzoli, W.~B. Messaoud, M.~Amberg, F.~Giraud, B.~Lemaire-Semail, and M.-A.
  Bueno, ``Physical and perceptual independence of ultrasonic vibration and
  electrovibration for friction modulation,'' \emph{IEEE transactions on
  haptics}, vol.~8, no.~2, pp. 235--239, 2015.

\bibitem{liu2020tri}
\BIBentryALTinterwordspacing
G.~Liu, C.~Zhang, and X.~Sun, ``Tri-modal tactile display and its application
  into tactile perception of visualized surfaces,'' \emph{IEEE Transactions on
  Haptics}, 2020. [Online]. Available:
  \url{https://ieeexplore.ieee.org/document/9034143}
\BIBentrySTDinterwordspacing

\bibitem{Emgin2015_HardwareDemo}
S.~Emgin, E.~Ege, O.~Birer, and C.~Basdogan, ``Localized multi-finger
  electrostatic haptic display,'' in \emph{Hardware Demonstration, 2015 IEEE
  World Haptics Conference (WHC)}, 2015.

\bibitem{haga2015electrostatic}
H.~Haga, K.~Yoshinaga, J.~Yanase, D.~Sugimoto, K.~Takatori, and H.~Asada,
  ``Electrostatic tactile display using beat phenomenon for stimulus
  localization,'' \emph{IEICE Transactions on Electronics}, vol.~98, no.~11,
  pp. 1008--1014, 2015.

\bibitem{biswas2019emerging}
S.~Biswas and Y.~Visell, ``Emerging material technologies for haptics,''
  \emph{Advanced Materials Technologies}, vol.~4, no.~4, p. 1900042, 2019.

\end{thebibliography}

% biography section
% 
% If you have an EPS/PDF photo (graphicx package needed) extra braces are
% needed around the contents of the optional argument to biography to prevent
% the LaTeX parser from getting confused when it sees the complicated
% \includegraphics command within an optional argument. (You could create
% your own custom macro containing the \includegraphics command to make things
% simpler here.)
%\begin{IEEEbiography}[{\includegraphics[width=1in,height=1.25in,clip,keepaspectratio]{mshell}}]{Michael Shell}
% or if you just want to reserve a space for a photo:
\begin{IEEEbiography}[{\includegraphics[width=1in,height=1.25in,clip,keepaspectratio]{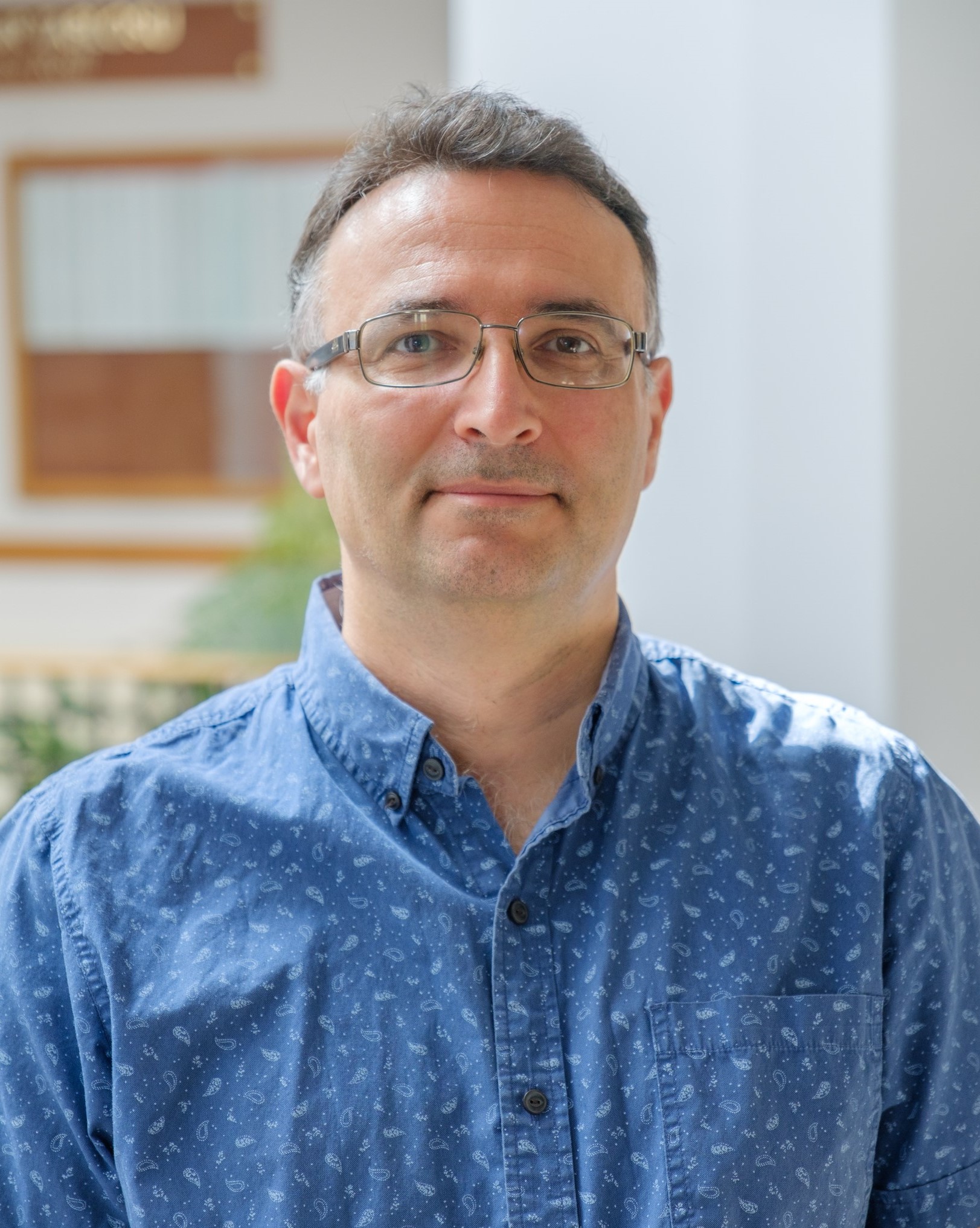}}]{Cagatay Basdogan}
received the Ph.D. degree in mechanical engineering from Southern Methodist University, in 1994. He is a faculty member in the mechanical engineering
and computational sciences and engineering
programs at Koc University, Istanbul, Turkey.
He is also the director of the Robotics and
Mechatronics Laboratory, Koc University. Before joining Koc University, he worked at NASAJPL/Caltech, MIT, and Northwestern University
Research Park. His research interests include
haptic interfaces, robotics, mechatronics, biomechanics, medical simulation, computer graphics, and multi–modal virtual environments. He
is currently the associate editor in chief of the IEEE Transactions on
Haptics and serves on the editorial boards of the IEEE Transactions on
Mechatronics, Presence: Teleoperators and Virtual Environments, and
Computer Animation and Virtual Worlds journals. In addition to serving in program and organizational
committees of several haptics conferences, he chaired the IEEE World
Haptics Conference in 2011.
\end{IEEEbiography}
\begin{IEEEbiography}[{\includegraphics[width=1in,height=1.25in,clip,keepaspectratio]{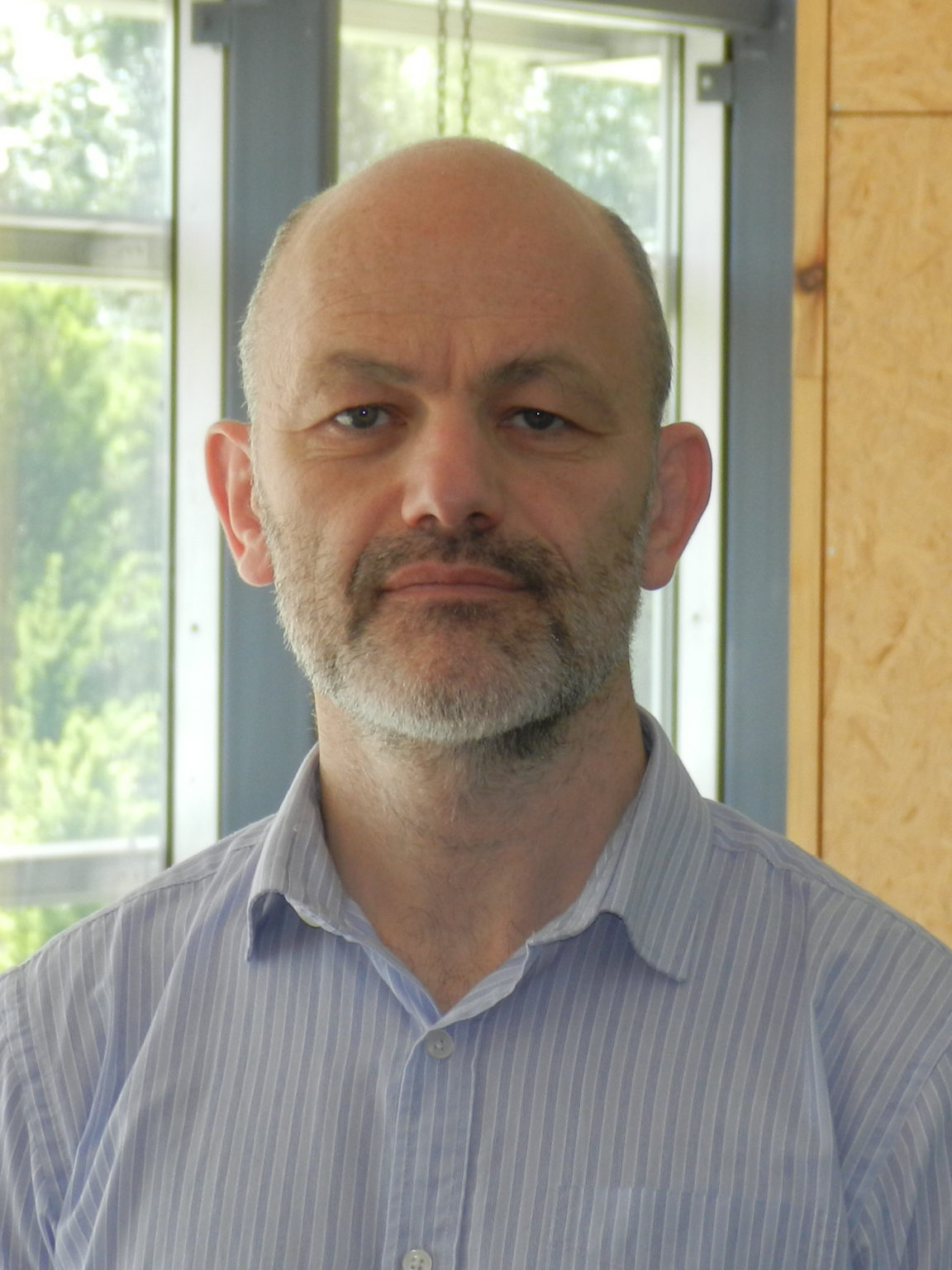}}]{Fr\'{e}d\'{e}ric Giraud} is an Associate Professor in electrical engineering at University of Lille and a member of the L2EP (Laboratory of Electrical Engineering and Power Electronics). His research interest is on the modelling and control of piezoelectric actuators, for Mechatronic systems and Haptic devices. He was a student at the Ecole Normale Sup\'{e}rieure de Cachan (1993-1996); he obtained a Master degree from the Polytechnical Institute of Toulouse (France) and a PhD from the University of Lille, both in electrical engineering. He is currently associate editor of the IEEE Transactions on Haptics.
\end{IEEEbiography}
\begin{IEEEbiography}[{\includegraphics[width=1in,height=1.25in,clip,keepaspectratio]{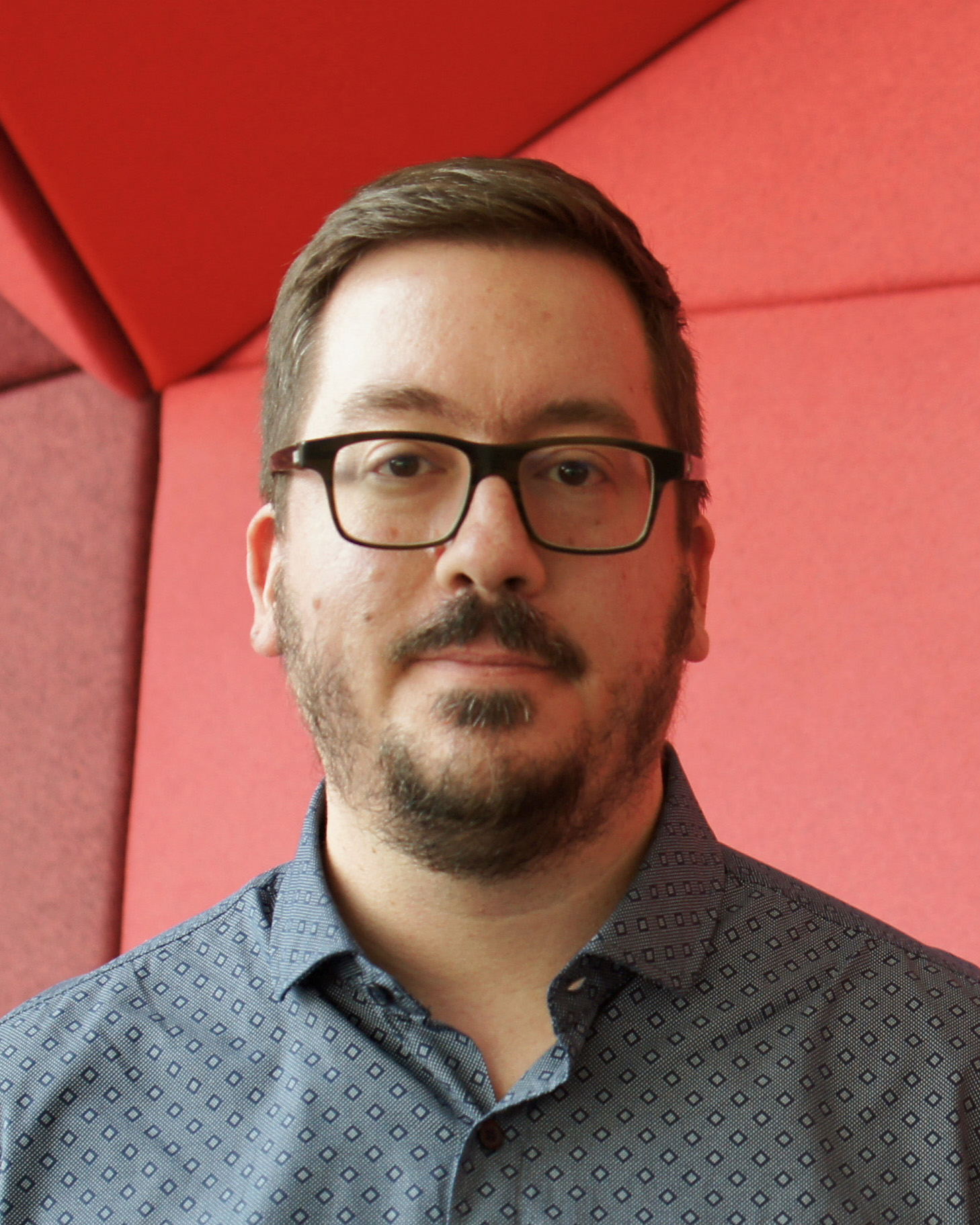}}]{Vincent Levesque} is an Associate Professor in the software and IT engineering department at École de technologie supérieure and the director of the Haptic User Experience (HUX) research group. \revc{He serves on the editorial board of the IEEE Robotics and Automation Letters and  is the general co-chair of the 2021 IEEE World Haptics Conference and the program chair of the 2020 Haptics Symposium. He served on the editorial board of the IEEE Transactions on Haptics from 2016 to 2019. He was a research scientist with Immersion Corp. from 2011 to 2016 and a Postdoctoral Fellow at the University of British Columbia from 2009 to 2011. He received a B.Eng. degree in Computer Engineering (2000), and M.Eng. (2003) and Ph.D. (2009) degrees in Electrical Engineering from McGill University. His research interests are at the intersection of haptic technologies and human-computer interaction, and include mobile and wearable haptics, tactile displays, and surface haptics. He is the recipient of several awards including the biannual 2019 Early Career Award of the IEEE Technical Committee for Haptics and Best Paper Awards at the 2012 Haptics Symposium and the 2011 ACM CHI Conference for his work on touch interaction with programmable friction.}
\end{IEEEbiography}
\vfill
\begin{IEEEbiography}[{\includegraphics[width=1in,height=1.25in,clip,keepaspectratio]{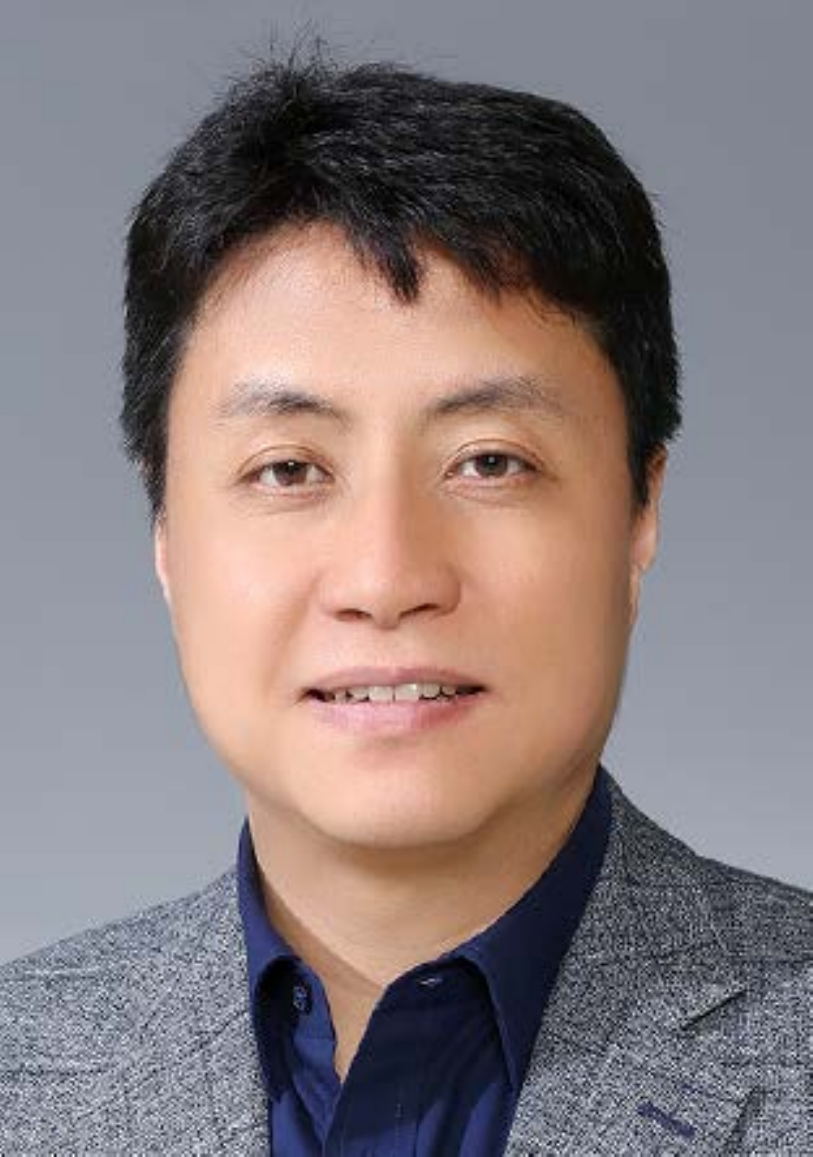}}]{Seungmoon Choi}
is a professor of Computer Science and Engineering at Pohang University of Science and Technology (POSTECH). He received the B.Sc. and M.Sc. degrees in Control and Instrumentation Engineering from Seoul National University in 1995 and 1997, respectively, and the Ph.D. degree in Electrical and Computer Engineering from Purdue University in 2003. He received a 2011 Early Career Award from IEEE Technical Committee on Haptics and several best paper awards from major international conferences. He was a co-chair of the IEEE Technical Committee on Haptics in 2009-2010. He serves/served on the editorial board of IEEE Transactions on Haptics, Presence, Virtual Reality, and IEEE Robotics and Automation Letters. He was the general co-chair of IEEE Haptics Symposium in 2014 and 2016 and the program chair of IEEE World Haptics 2015. His research interests lie on haptic rendering and perception, both in kinesthetic and tactile aspects. His basic research has been applied to mobile devices, automobiles, virtual prototyping, and motion-based remote controllers. He is a senior member of the IEEE.
\end{IEEEbiography}

% You can push biographies down or up by placing
% a \vfill before or after them. The appropriate
% use of \vfill depends on what kind of text is
% on the last page and whether or not the columns
% are being equalized.
%\vfill

% Can be used to pull up biographies so that the bottom of the last one
% is flush with the other column.

\vfill
%\newpage
%\input{img/Table_HM.tex}
% that's all folks
\end{document}